\documentclass{article}

\usepackage{graphicx}				
\usepackage{amsmath}				
\usepackage{amsfonts}
\usepackage[colorlinks=true,linkcolor=blue,anchorcolor=black,citecolor=blue,filecolor=black,menucolor=black,urlcolor=blue,breaklinks=true,pdfhighlight=/P,pdfmenubar=true,pdftoolbar=true,pdfpagelabels=true,pdfstartpage=1,pdfstartview=FitV,pdftitle={User Equilibrium Route Assignment for Microscopic Pedestrian Simulation},pdfsubject={Pedestrian, Crowd, Mass, Simulation, Dynamics, VISWALK, VISSIM},pdfauthor={Kretz},pdfcreator={tobias Kretz},pdfproducer={t kretz},pdfkeywords={Pedestrian, Crowd, Mass, Simulation, Dynamics, VISWALK}]{hyperref}	
\usepackage[numbers,sort&compress]{natbib}	
\usepackage{hypernat}				
\usepackage[american]{babel}		
\usepackage{simplemargins}
\setallmargins{1in}
\usepackage[american]{babel}		
\title{User Equilibrium Route Assignment for Microscopic Pedestrian Simulation}

\author{Tobias Kretz (corresponding author)$^1$, Karsten Lehmann$^{1,2}$, Ingmar Hofs{\"a}{\ss}$^1$ \\
$^1$: PTV Group, Haid-und-Neu-Stra{\ss}e 15, D-76131 Karlsruhe, Germany\\
\tt{\{Firstname.Lastname\}@ptvgroup.com} \\
phone: +49 721 9651 7280 \\
$^2$: init AG, K{\"a}ppelestra{\ss}e 4-6, D-76131 Karlsruhe, Germany
}%
\begin{document}
%

\maketitle

\begin{abstract}
For the simulation of pedestrians a method is introduced to find routing alternatives from any origin position to a given destination area in a given geometry composed of walking areas and obstacles. The method includes a parameter which sets a threshold for the approximate minimum size of obstacles to generate routing alternatives. The resulting data structure for navigation is constructed such that it does not introduce artifacts to the movement of simulated pedestrians and that locally pedestrians prefer to walk on the shortest path. The generated set of routes can be used with iterating static or dynamic assignment methods.
\end{abstract}

{\bf Keywords:} dynamic assignment, pedestrians, microsimulation




\section{Introduction}
A model of pedestrian dynamics in the simplest case has to include three separate elements: 1) a method to determine the main walking direction towards the destination, 2) a method to avoid collisions respectively to resolve conflicts between pedestrians and 3) a method to avoid bumping into walls and even more important to avoid passing through walls.

Issues 2) and 3) have been addressed in various ways, with forces \cite{Helbing1995, Helbing2000, Yu2005, Johansson2007, Chraibi2010}, in velocity space \cite{fiorini1998motion, vandenberg2008reciprocal,vandenberg2011reciprocal} or directly in space \cite{Burstedde2001, kirchner2002simulation, Kluepfel2003, Nishinari2004, Kretz2006d}.

These {\em operational} models of pedestrian dynamics usually also address issue 1). They might apply different methods to solve it, but normally they assume that it is the direction of the shortest path (fewest meters) to destination which pedestrians are heading for. Quote from \cite{Helbing1995}: {\em ``He/She wants to reach a certain destination $\vec{r}^0_{\alpha}$ as comfortable as possible. Therefore he/she normally takes a way without detours, i.e., the shortest possible way.''} See subsection \ref{sec:Navigation} for a description of two very common ways to implement this principle.

This idea may be correct or at least a good approximation in situations with a very low pedestrian density. However, at higher densities and in particular with locally varying densities collision avoidance and conflict resolution imply discomfort and time delays. If collision avoidance and conflict resolution are required over proportionally often along the shortest path the more comfortable and/or the quicker path to destination may be along some detour.

Obviously when pedestrian density is moderate to high only few, if any, pedestrians can really walk on the shortest path. Everyone else has to do some additional meters. In models of pedestrian dynamics as referenced above these extra meters result from collision avoidance or conflict resolution which in the simplest case is just a hard-core exclusion. However, such deviations from the shortest path are not detours in the sense of the previous paragraph. Detours in the sense of the previous paragraph are paths which result from pedestrians trying to spread out to reduce density, increase comfort, and reduce travel times. For the matter of models of pedestrian dynamics: a simplest case model of pedestrian dynamics is extended by an element which makes {\em pedestrians avoid having to avoid collisions} or {\em prevents pedestrians having to resolve conflicts}.

There are various approaches to address this problem. Some of these will be discussed in section \ref{sec:previous} on previous work. 

The approach adopted in this work is to create a relatively small set of routing alternatives between origin and destination. With the help of intermediate destinations each pedestrian is made to walk along one of the routes. While approaching an intermediate destination a pedestrian can be controlled with one of the operational models mentioned above. In particular each pedestrian can walk into the direction of the shortest path. Navigation along the shortest path is simple and computationally cheap because the map of distance to (intermediate) destination is a constant contrary to the metric of estimated travel time to (intermediate) destination \cite{kretz2009a, Kretz2009c, Kretz2010c, Kretz2010x}. The detour compared to the shortest path between origin and destination only results as a consequence of sequentially approaching a number of intermediate destinations. Therefore the proposed approach aims at combining the efficient navigation along the shortest path with pedestrians who are able to walk detours between origin area and destination area by passing a sequence of intermediate destinations.

Two routes between the same origin and destination can be different, but are still not perceived as actual routing alternatives by a pedestrian or car driver because they are not sufficiently distinct \cite{chen2007reliable,abraham2010alternative,dees2010defining,bader2011alternative,luxen2012candidate,abraham2013alternative}. An example from road traffic can easily exemplify this: let's say the best or at least a good route -- judged on travel time or distance or  some other factor or a combination of many factors -- from Paris to Berlin passes by Cologne. A na{\"i}ve computation of the second best route would probably bring up a solution where some 100 meters of the highway are replaced by the parallel section via a highway parking place. However, the user of a navigation device would not actually {\em perceive} this as a real (substantial, relevant) alternative to the first solution, but rather a route maybe via Antwerp or Frankfurt although both of these routes have a higher expected travel time than the route via the highway parking place near Cologne. The routing alternatives computed by the method proposed in this work are mutually distinct not only in the sense that they cannot be transformed one into the other with a homotopic operation, but mutually distinct by a configurable minimum degree -- they are substantially different, respectively pose relevant alternatives.

Last but not least the intermediate destinations are computed geometrically such that locally they do not change pedestrian behavior in particular not introduce unnatural artifacts into the movement dynamics. I.e. someone observing an animated visualization of the simulation could not tell the presence of an intermediate destination. 

In the remainder first a motivation for the proposed method is given. The consequent section contains a discussion of preceding relevant work. This is followed by the section which defines the method. Finally the method is applied to an example. In the appendix we give definitions for the phrases we use in this work.

A note to readers who are not entirely familiar with all topics which are touched in this paper: it may be easier to read first the section giving an example (\ref{sec:example}) as it should give an intuitive understanding of the properties of the simulation system as well as the results of the method which is proposed and only then proceed here.

\section{Motivation: Computing User-Equilibria}
Finding and evaluating the user equilibrium network load for road networks (traffic assignment) is one of the core tasks of traffic planning. After Wardrop had formulated the principles nowadays named after him \cite{wardrop1952some} various algorithms to solve this problem have been developed over the years \cite{Beckmann1956,leblanc1975efficient,bar2002origin,gentile2009linear}. 

The main hindrance for the application of these algorithms with pedestrian traffic is that all these algorithms are formulated for application with discrete networks, i.e. they compute flows on a node-edge graph. Pedestrians on the contrary can and do move continuously in two spatial dimensions. Their walking area can have holes (obstacles), but this does not change the fact that even when loops are excluded there are infinitely many possible paths for a pedestrian to walk from an origin to a destination. 

There are two ways to face the resulting challenge: either one develops an entirely different assignment algorithm for pedestrian traffic -- this was done for example in \cite{hoogendoorn2004pedestrian,hoogendoorn2004dynamic} -- or one extracts from the pedestrian walking area in a well-grounded suitable way a graph structure which can be used with existing assignment methods. The routes computed with the method presented in this contribution can be seen to be such a graph structure.

A user equilibrium can only be found in a mathematically thorough way with an iterative method. Any non-iterative (one-shot) method can only come close to an equilibrium by chance and will work better in one scenario and worse in a different one. ``Iterative'' means that the results of one iteration step are used to determine the route choice ratios (i.e. the loads for the routes) in the next iteration step\footnote{An iteration step can comprise of one or more simulation runs. In the latter case the results are aggregated over the various runs to get the foundation for the next iteration step.}. ``Non-iterative'' means that route decision ratios are dependent on what happens or has happened in the same simulation run.

The resulting graph of the method proposed in this paper cannot only be used with such iterated methods that aim for the user equilibrium, but also with other route choice problems and methods. However, iterated assignment aiming to find a user equilibrium can be seen as the ``core business'' of the method presented in this paper\footnote{At least this is currently the authors' opinion. Time might show that the routes computed with the methods of this article are more often used outside the scope of iterated assignment. This would not limit the relevance of the method at all. Just for the moment we are excited to have the foundation to combine long standing models of pedestrian dynamics with iterated assignment.}. The reason for this is that there are more non-iterative (``one-shot'') approaches for route choice. The section on existing work will discuss some of these and clarify this claim. With more alternative options available the method presented in this work appears to be less essential. For that reason in the following three paragraphs we motivate the formulation of the routing graph generation method of this paper by emphasizing the relevance of the user equilibrium in pedestrian traffic.

So, how relevant is user equilibrium in pedestrian traffic? After all on vehicular road networks a near-equilibrium state occurs day by day because of the commuters' workday by workday experience with a traffic system which is near or beyond its capacity. Are there comparable situations (repeated near-system-capacity demand which makes pedestrians accept to walk considerable detours to reduce walking time) for pedestrians as well? It is probably safe to say that there are less such situations and that they can be less clearly spotted compared to the daily vehicular rush hour. However, they can exist for example in large metropolitan stations \cite{Heuvel2014} and will more often occur there in the future considered the contemporary global urbanization dynamics \cite{UNO2012world}. 

Even in situations where one cannot assume that a user equilibrium is realized because there is no repeated experience with the situation there can be a clear benefit in knowing the user equilibrium. Take as example emergency evacuations. Emergency evacuations are a highly relevant application case for pedestrian simulations and emergency evacuations are the exact opposite of commuting regarding experience: even if emergency evacuation trainings are held frequently one can say that a real emergency evacuation usually is the first time experience for all occupants at least concerning the particular building. For usually the user equilibrium will be the more efficient solution than the distribution which emerges naturally in an emergency evacuation, a planner can take measures to shift the system toward the user equilibrium. In emergency evacuation planning this could for example be done with additional or different route signage. \footnote{One might ask if it was more desirable then to know the system optimum distribution (for which the routes computed with the method of this article could also serve as a base). However, particularly for emergency evacuation planning attempting to implement with signage the system-optimal distribution rather than the user equilibrium raises an ethical question: is it allowed to send a few people on paths which are suboptimal for them to benefit many others? The second argument against going for the system optimum is a question of its relevance: the experience in vehicular road traffic is that in realistic networks often there is not much of a difference between user equilibrium and system optimum \cite{leblanc1984comparison}. Despite the existence of Braess' Paradox in vehicular road traffic \cite{braess1968paradoxon} and despite the fact that one can construct examples with counter-intuitive effects for pedestrian traffic as well \cite{Kretz2011counterflow} one can suspect that for realistic examples the user equilibrium can serve as a good solution and thus as benchmark to compare other solutions with.}

As a third argument in favor of the relevance of user equilibrium assignment for pedestrian traffic simulations one can hypothesize that pedestrians at least in certain situations  have an intuitive understanding of how a pedestrian flow situation will evolve even if they are faced with the particular situation for the first time and that they will distribute close to the user equilibrium. 

Of course pedestrians do not in all situations distribute according to the user equilibrium. Pedestrian route choice can depend on many other factors, including the availability and distribution of landmarks as well as social aspects \cite{graessle2011example}. Universality, however, is not required that a method to compute it is relevant. It is sufficient if these situations exist and planners have an understanding which situations these are.

\section{Existing Work} \label{sec:previous}
\subsection{Models of Operational Pedestrian Dynamics}
There is a great number of models of operational pedestrian dynamics which route pedestrians on the spatially shortest path. These models fall into a number of categories. For example so-called Cellular-automata models are manifestly formulated in discrete space and discrete time \cite{Kluepfel2000,Burstedde2001,Kretz2006d}. Different to these force-based models as the Social Force Model \cite{Helbing1995,Johansson2007} and most other force-based models \cite{Yu2005,Chraibi2010} are formulated in discrete time and discrete space. Usually their temporal aspect can be discretized or has to be discretized as the only way to solve the system equations is a numerical integration \cite{seyfried2006basics}. For an overview of these and other operational models of pedestrian dynamics see for example \cite{Schadschneider2009b, Johansson2012a}. These models are interesting here in so far as -- according to the motivation stated above -- they lack an important element which is supplied with the method proposed in this paper. The sheer number of these models underlines the relevance of our method yet on the other side makes it impossible to discuss or even just mention all of them here. Instead we will shortly discuss some models and methods which address the same issue as our method does.

\subsection{Models of Pedestrian Dynamics which Include Elaborate Route Planning Methods}
To the authors' knowledge so far there has not been proposed another general method that analyzes and computes pedestrian routing alternatives in a thorough way such that it can be used with existing methods for iterated assignment of a microscopic pedestrian simulation without introducing artifacts into the movement of pedestrians. However, if the stated properties are contrasted with their opposite -- iterated vs. one-shot, microscopic vs. macroscopic, discrete route vs. continuous space, and time interval vs. continuous time -- then there has been a number of contributions which aim in the same direction of making simulated pedestrians avoid having to avoid conflicts.

An iterated assignment in continuous space and time with a macroscopic approach has been proposed Hoogendoorn and Bovy in \cite{hoogendoorn2004dynamic}. The paper -- together with \cite{hoogendoorn2004pedestrian} -- also gives a good problem formulation and overview of the fundamentals of pedestrian assignment. Different to our method and as usual for continuous space approaches this method does not make route choices explicit in the sense that it generates routes as identifiable data objects. One can assume that such a method is computationally very demanding as it considers infinitely many routes instead of singling out relevant route choices to work with these. 

Earlier Hughes proposed a one-shot macroscopic method \cite{hughes2002continuum,huang2009revisiting}. As it is different from our approach in two aspects (one-shot {\em and} macroscopic) we will not discuss it further here.

Kemloh Wagoum et al. proposed a dynamic one-shot route choice method with discrete routes which distinguishes between locally and globally optimal routes and between shortest and quickest path \cite{kemloh2012modeling}. A node-link network structure is extracted from the geometry where rooms are nodes and doors or corridors are links. Pedestrians with a (locally) quickest routing strategy observe other pedestrians in the same node (room) and their current speed. Based on this they make an estimation on what is the quickest route. The focus of the work clearly is set on modeling the ``internal state'' of a pedestrian -- if he or she is in a hurry or not or patient or not -- and thus applies a shortest or a quickest path strategy. It is furthermore modeled in an elaborate way how he or she selects a reference pedestrian and draws conclusions from this pedestrian's current walking speed. There is on the contrary no discussion of the semantics of a walking infrastructure. It is implicitly assumed that rooms and corridors (or doors) can always be identified easily and uniquely, hereby assuming that doors or corridors are the bottlenecks, i.e. the capacity limiting elements and that rooms are spill-back and jamming areas, which have no effects on capacity. These assumptions hold for functional building infrastructure as for example office buildings. Station halls, airport terminals and urban public spaces as examples on the contrary usually are comprised of a mixture of obstacles and walking spaces which does not allow such a clear distinction between room and corridor respectively node and link. As a consequence the possibility to distinguish -- in the sense of the algorithm -- between global and local gets lost. The algorithm proposed by Kemloh et al. may be applicable in such a scenario or at least a share of such scenarios. Then, however, the node-link network would have to be created specifically for this application in a particular scenario. The paper does not give a general method for its construction, i.e. the focus of this paper is not a method to produce routing alternatives, but on the method to decide which route to take.

Another approach is to base navigation on a visibility graph \cite{deBerg1997} or a graph derived from it \cite{kneidl2012generation} and either use -- in a one-shot approach -- the pedestrian densities in the vicinity of the links to estimate current travel times and thus navigate pedestrians on the instantaneously estimated quickest route \cite{bluemel2008,hocker2010graph} or the travel times on the links in the previous time interval of the same simulation run or the travel times on the links in a previous iteration in an iterative approach. Problems arise from the visibility graph only offering navigation points. This inescapably leads to problems which will be discussed below. Furthermore a visibility graph also includes route alternatives generated by very small obstacles like litter bins which can be handled by the operational part of a simulation model.

A continuous space, continuous-time, one-shot method was introduced by Treuille et al. in \cite{treuille2006continuum}. Crowd movement here is determined by the computation of an optimal path under consideration of distance, travel time (resp. speed), and discomfort, which results in one single ``dynamic potential'' respectively its differential form the ``unit cost field''. While pedestrians are represented microscopically the interaction between them is exclusively transferred via the common dynamic potential i.e. in a macroscopic manner. The model therefore can arguably be referred to as a mesoscopic model. This method as well does not generate routes as identifiable data objects.

Based on the principle of least effort referring to energy consumption Guy et al. proposed a related continuous-time, one-shot method in which a strong congestion avoidance behavior emerges \cite{guy2010pledestrians}. In this model it is the current average speed in the vicinity of the links of a precomputed node-link roadmap\footnote{Only few details on the nature of the roadmap are given in the paper.} which determines the expected energy to traverse a link and thus route choice. One could consider this method as ``almost continuous space'' as the pedestrians are moving in this way, while the route choice determining effects are transmitted via a node-link graph. 

Another continuous space, (almost) continuous time, one-shot method for use with microscopic simulation has been introduced by Kretz et al. in \cite{Kretz2011e,Kretz2013c}. In this approach a map of estimated remaining travel time to destination (called ``dynamic potential'') of which the pedestrians use the negative gradients as their desired walking direction to walk into the direction of the quickest path. This desired walking direction then is used in a variant of the Social Force Model \cite{Johansson2007}. This implies interaction between identified individuals and therefore that the whole model is truly microscopic. The dynamic potential is computed in instantaneous time, i.e. for locations ahead of a pedestrian the current conditions are used as input for the computation and not (estimated) future conditions of the time when the pedestrian (potentially) will be there. This is not necessarily a drawback for the degree of realism, as real pedestrians also have only a very limited capacity to guess future conditions beyond a few seconds unless the situation can be recognized to be about steady-state in which the dynamic potential approach also works fine. While the simulation results appear realistic, the approach does not indicate if the evolution of the simulation is close to an equilibrium solution. 

Related to the continuous space, one-shot methods is an idea by van Toll et al. to explicitly measure density on polygon-shaped, discrete, non-overlapping areas and compute from these densities expected travel times along a navigation map \cite{vantoll2012realistic}.

Other related work includes \cite{abdelghany2012dynamic} and \cite{Gao2013}.

\subsection{Navigation Methods} \label{sec:Navigation}
There is a large amount of literature on navigation methods for the simulation of pedestrians or control of robots \cite{Khatib1986}. Navigation means to guide a simulated pedestrian or a robot to a given destination while considering obstacles. A simulated pedestrian or a robot should neither bump into obstacles nor get trapped in a dead-end and not find out anymore. There are mainly two methods to solve this problem: with sparse {\em navigation graphs} or with {\em distance maps} also called {\em static potentials}. We understand navigation here as the element of a method which sets the basic direction of movement. There can be other elements in a model of pedestrian dynamics which modify this basic direction. For example for the Social Force Model we think of the driving force term as incorporating the navigation aspect while the forces from other pedestrians are a refinement which one could call ``sub-navigational''.

\begin{figure}[htbp]
  \center
	\includegraphics[width=0.305\textwidth]{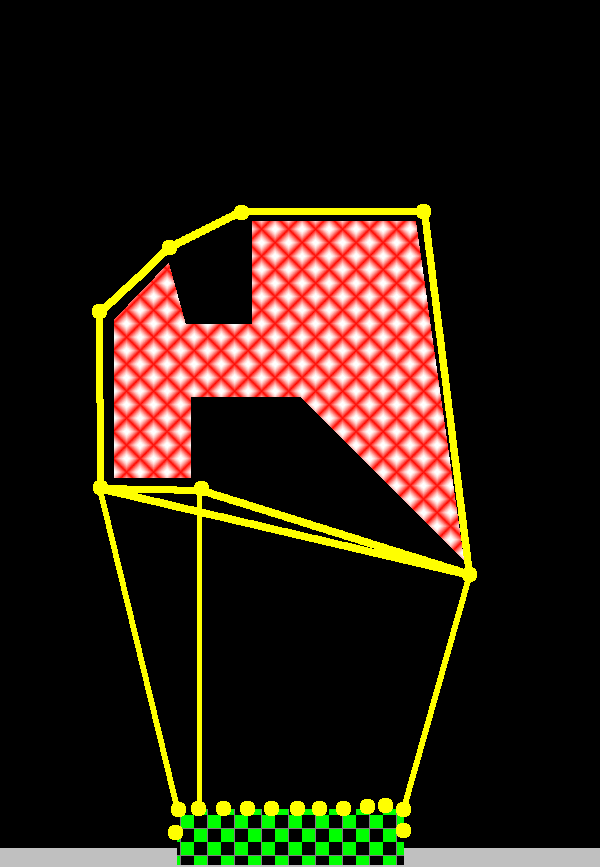} \hspace{12pt}
	\includegraphics[width=0.305\textwidth]{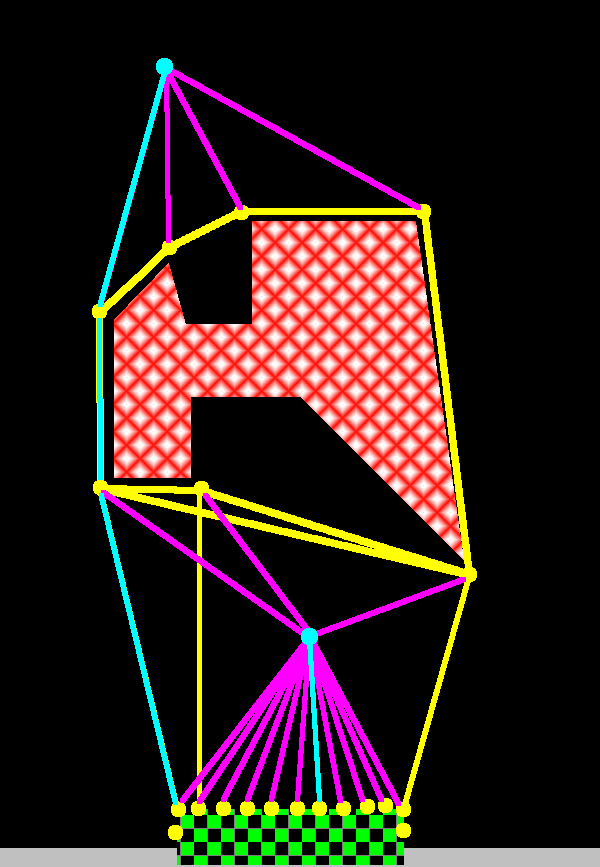} \\
	\caption{Left: Example of a visibility graph, resp. navigation graph. Walking areas of pedestrians are shown black, an obstacle in the middle is marked with diagonal red stripes on white ground. The destination at the lower edge is shown as green and black checkerboard pattern. The navigation graph and the navigation nodes are marked yellow. Note that on the destination area a number of navigation nodes have to be put such that pedestrians for which the destination area is visible from the start walk the about the shortest path to the destination area and not a longer way to one of its corners. The right image shows in addition two possible origin positions (cyan dots) and links to the nodes which are visible (magenta and cyan). The shortest navigation paths to the destination area are marked cyan.}
	\label{fig:navigation-graph}
\end{figure}

The idea of navigation graphs is to have a small set of vertices -- navigation points -- of which at each time each pedestrian has one where (s)he is heading for. The vertices are connected to a graph in a way that a link exists, if two vertices are mutually visible, i.e. there is no obstacle in the direct line of sight. As visibility is the guiding principle in the construction of such a graph it is usually called a {\em visibility graph} \cite{deBerg1997}\footnote{Another variant to construct a basic graph is to compute the (generalized) Voronoi diagram of the obstacles \cite{karamouzas2009indicative}. An elaborate variant introduced in \cite{pettre2005navigation} for example largely reduces the problem of artificial bottlenecks, but still navigation elements (in said paper called`{\em navigation graph edges}) generally introduce artifacts to the local motion as illustrated in section \ref{sec:illustration}.}. Further construction principles are that the graph should have as few nodes as possible and that navigating a single pedestrian from node to node allows navigation on the shortest path\footnote{A navigation graph could also be applied with a different paradigm for example that the walking direction should have a minimum angle with some landmark. However, most of the time walking distance is the decisive criterion.} to the destination for any origin position. Once a navigation graph is computed the shortest paths on it can easily be computed using for example Dijkstra's algorithm \cite{Dijkstra1959} Figure \ref{fig:navigation-graph} shows a simple example of a navigation graph which leads the shortest way around one single obstacle to a destination.

Navigation graphs are an intuitive and almost self-suggesting approach for navigation in simulation of pedestrians and control of robots \cite{Latombe1991}. The method works well to navigate individuals. Problems occur when it is used to navigate groups -- crowds or swarms.  As swarms of robots are rare -- although this has begun to change  \cite{marcolino2008no} -- the method usually works well in the field of robotics. To navigate crowds of pedestrians (or swarms of robots) the method can become problematic. Having the navigation graph and the correct sequence of nodes to the destination is not sufficient. At each time step of a simulation one has to select one of the nodes as current destination node. 

The simplest way is to define an {\em arrival distance} to which a pedestrian has to get close to a navigation node and if the pedestrian has come that close or closer in later simulation time steps the next node is used as destination node. If only a single pedestrian is moving in the vicinity of a navigation node the arrival distance can be set at a small value (about a body diameter). It even has to be set such small as a larger arrival distance would make the pedestrian proceed too early with the subsequent node and eventually make him bump into a wall, even get trapped in a dead-end. If on the contrary there is a large group of pedestrians heading for a navigation node in reality not all pedestrians pass by the node really close. If one has set a small arrival distance a navigation node poses an artificial bottleneck: one after the other pedestrian of the group has to pass by close to the navigation point to be able to proceed -- compare the discussion in \cite{koster2013avoiding} and for example figure 6 of \cite{kneidl2012generation}. One might attempt to fix the problem by adjusting for each navigation node individually the arrival distance depending on the surrounding pedestrian density, but this is a rather complicated and computation intensive fix which takes away much attractiveness from the original approach drawn from its simplicity.

\begin{figure}[htbp]
  \center
	\includegraphics[width=0.612\textwidth]{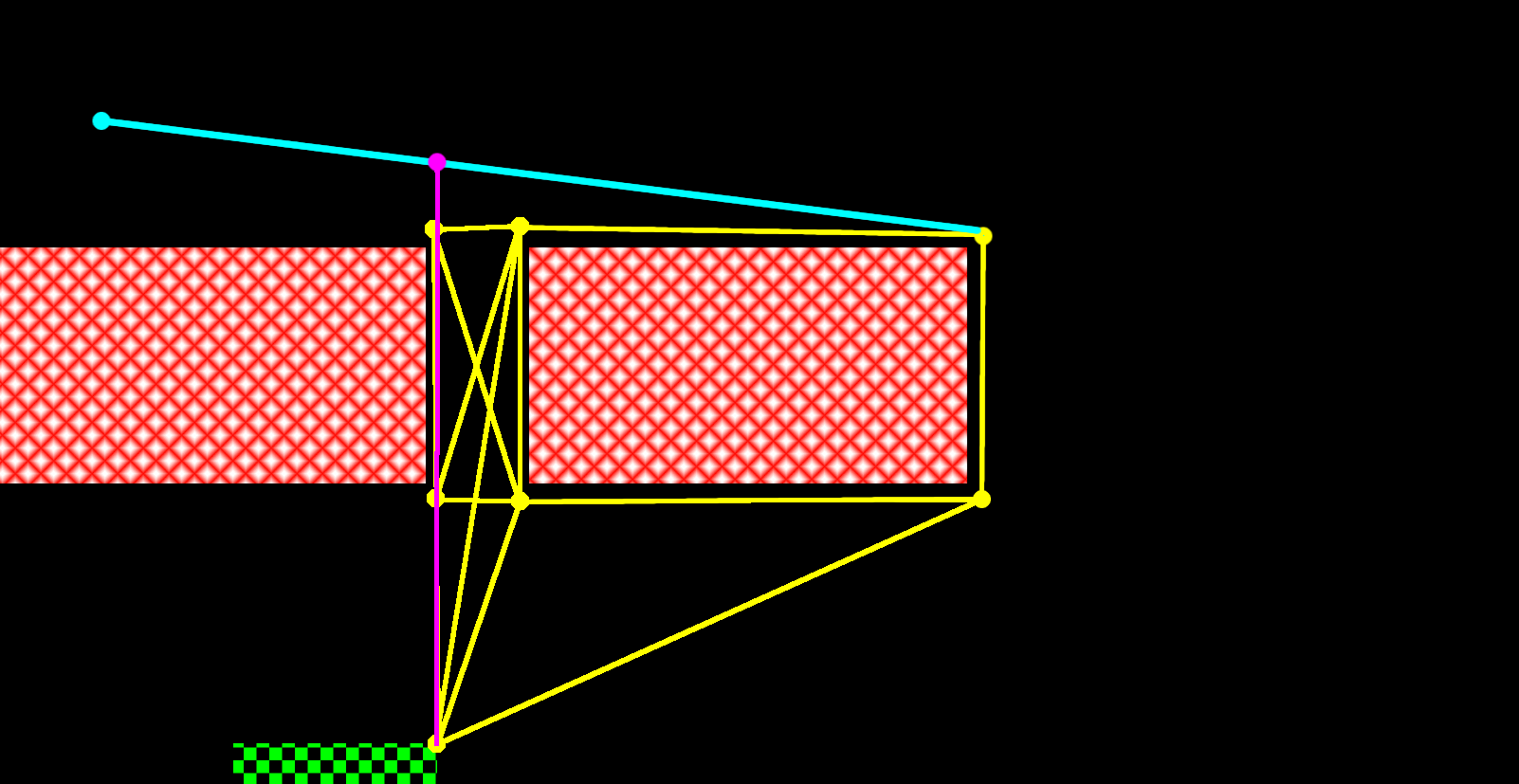} 
	\caption{This figure shows the origin position of a pedestrian (cyan dot) who is assigned to walk a detour on a path which has a higher capacity compared to the shortest possible path (which is what the facing contribution is about). Assume a crowded scenario where some pedestrians have to walk the detour as the shortest path has not enough capacity. The first navigation node where a detouring pedestrian has to head for is the one which is linked to the origin position with the cyan link. Then the pedestrian at some later time during the simulation will be located at the magenta spot. If the rule ``move toward that visible node which is closest to the destination'' is applied the pedestrian will cancel his detour and walk through the narrow corridor, i.e. the shortest path instead of the intended detour. This implies that for the purpose of this contribution one would have to steadily monitor if the rule can be applied or not and decide how else to choose a navigation node if not according to said rule.}
	\label{fig:failing-navigation-graph}
\end{figure}

A second possible way to select a navigation point is to make pedestrians always head for that navigation node which is a) visible from the pedestrians current position and b) closest to the destination area. This immediately fixes the problem of pedestrians possibly getting stuck or bumping into a wall at the cost of having to permanently do costly visibility computation throughout the simulation. For the purpose of this contribution this second way to choose navigation node is ruled out for another reason which is illustrated and explained in figure \ref{fig:failing-navigation-graph}.

As a consequence of these difficulties the approach of a navigation graph is discarded as a whole for the purpose of this contribution. Distance maps respectively potential fields are chosen instead as base method.

\begin{figure}[htbp]
  \center
	\includegraphics[width=0.305\textwidth]{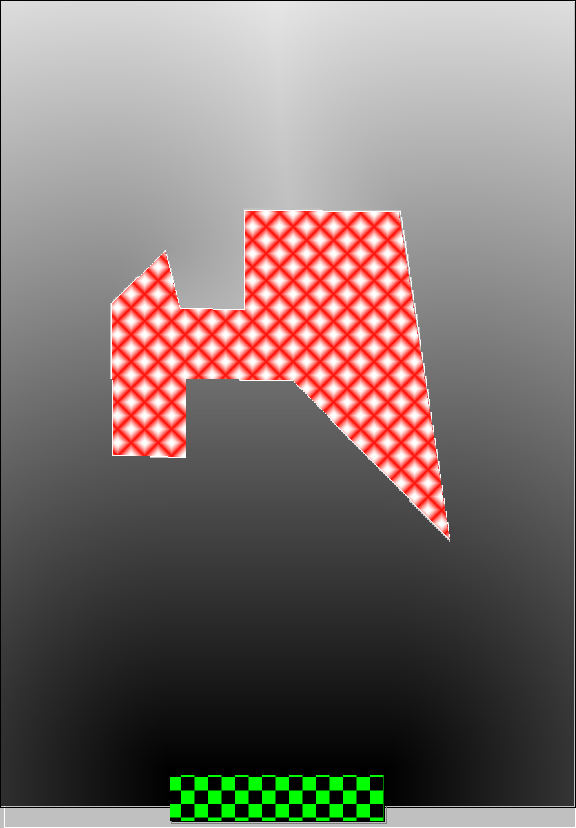} 
	\includegraphics[width=0.305\textwidth]{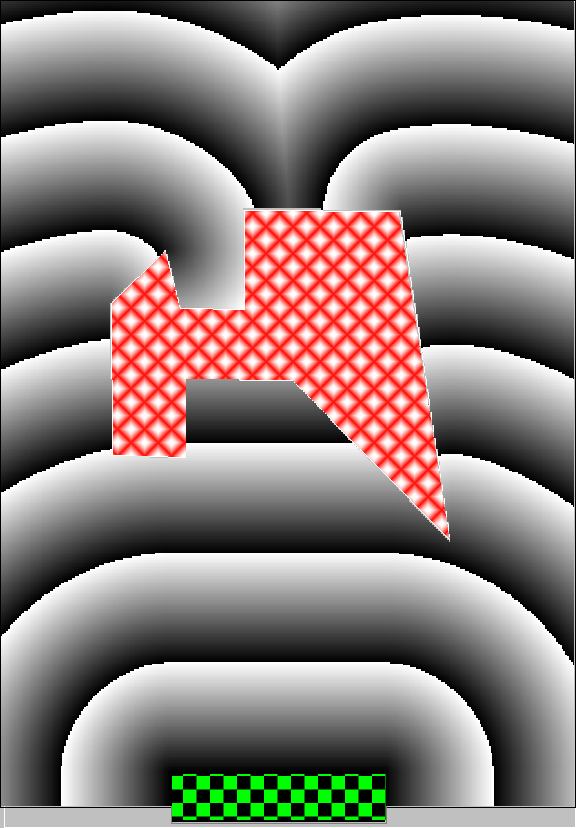} 	
	\caption{The left figure shows a distance map with the brightness of a spot directly proportional to the value of the grid point, i.e. the distance of that spot to the destination area. The destination area is shown as a green-black checkerboard, the obstacle is marked with diagonal red lines on white ground. In the right figure the brightness is computed modulo to some distance value $d$.}
	\label{fig:potential}
\end{figure}

The method of distance maps or static potentials has been popular for modeling pedestrian dynamics in particular in use with cellular automata models \cite{Burstedde2001,kirchner2002simulation,Kluepfel2003,Nishinari2004,Kretz2006f,yanagisawa2007mean,kirik2007intelligent,dudek2007application,ali2008floor}. The reason for this is probably that the data structure of the distance map fits well to the spatial representation of cellular automata models: a regular usually rectangular grid. Figure \ref{fig:potential} visualizes an example of a distance map.

\begin{figure}[htbp]
  \center
	\includegraphics[width=0.612\textwidth]{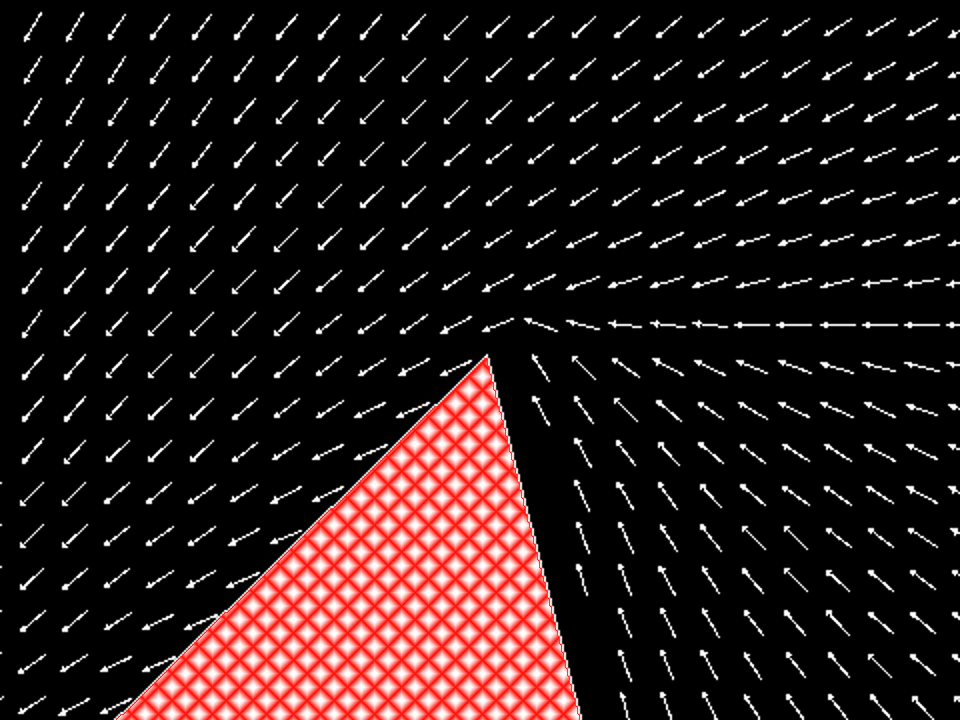} 
	\caption{Field of negative gradients (white). This figure shows an excerpt of the geometry of figure \ref{fig:potential}. Shown is the area around the upper left corner of the obstacle. It is a general mathematical property of gradients that they are orthogonal to lines of equal value (equi-potential lines) of the field from which they are derived. Therefore the negative gradients give the direction which for a given step distance most reduces the distance to destination. Basically following the negative gradients pedestrians are also guided around the obstacle.}
	\label{fig:gradients}
\end{figure}

In cellular automata models the distance maps are usually used such that pedestrians move with higher probability to a cell which is closer to the destination. In other modeling approaches which do not directly operate with space, but rather with velocity or acceleration as for example the force-based models do\footnote{An example for a model where the gradients are used for the velocity instead of he acceleration (as in force-based models) is discussed in \cite{dietrich2014navigation}}, pedestrians follow the negative gradient of the distance map at their current position. ``Follow'' here means that this direction is used as preferred or desired or base direction and may be subject to modifications as consequence of influence of for example other pedestrians or limited acceleration ability. Figure \ref{fig:gradients} shows an example for the field of gradients.

For reasons of simplicity and computation speed some models of pedestrian dynamics compute the distance map using a simple flood fill method: at the destination area all grid values are set to zero. Then a ``flood'' is driven outward simply either moving only over common edges or over common edges and common corners from cell to cell. In each of these steps a counting parameter is increased by one. The value of this counting parameter is set as grid cell value when the flood front passes the grid cell. The resulting metric is not Euclidean, but either Manhattan metric (if in the flooding process only common edges are passed) or checkerboard metric (if also common corners are passed). A detailed discussion of these methods and variants can be found in \cite{Kretz2010a}.

The most common method to compute a nearly Euclidean distance map is the one of \cite{Kimmel1998} for which meanwhile also variants exist which promise lower computation times under certain conditions \cite{Zhao2005,Jeong2007} and there has been proposed a flood fill-like method which also produces a Euclidean metric \cite{Schultz2010a}.

\section{Problem Illustration} \label{sec:illustration}
\begin{figure}[htbp]
  \center
	\includegraphics[width=0.612\textwidth]{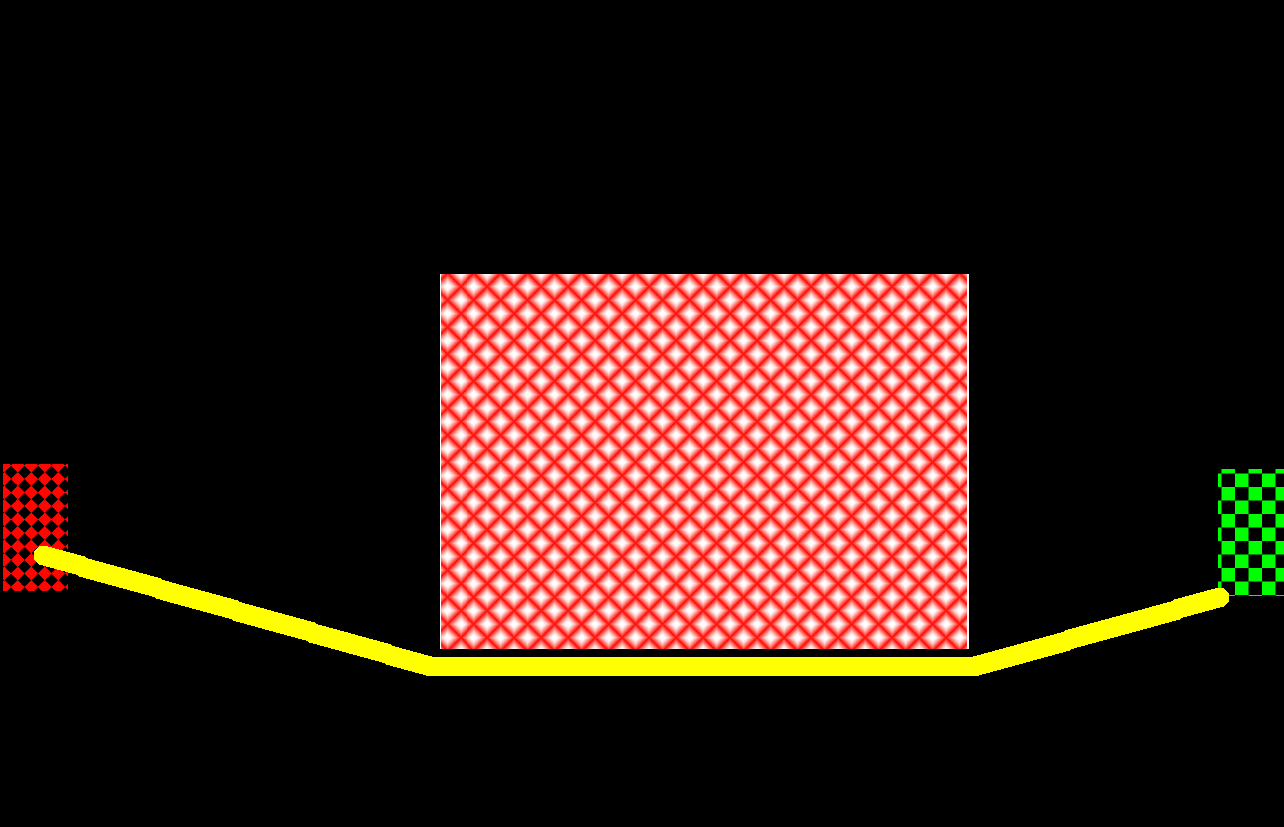} 
	\caption{Walking area (black), obstacle (diagonal red lines on white ground), origin area to the left (red and black diagonal checkerboard pattern) and destination area to the right (green and black checkerboard pattern). The yellow line shows the path a pedestrian (set into the simulation at some arbitrary coordinate on the origin area) would follow if a distance map is used to determine his basic direction. The route data simply would be some information (e.g. an area ID) identifying the destination area.}
	\label{fig:Example1010}
\end{figure}

\begin{figure}[htbp]
  \center
	\includegraphics[width=0.612\textwidth]{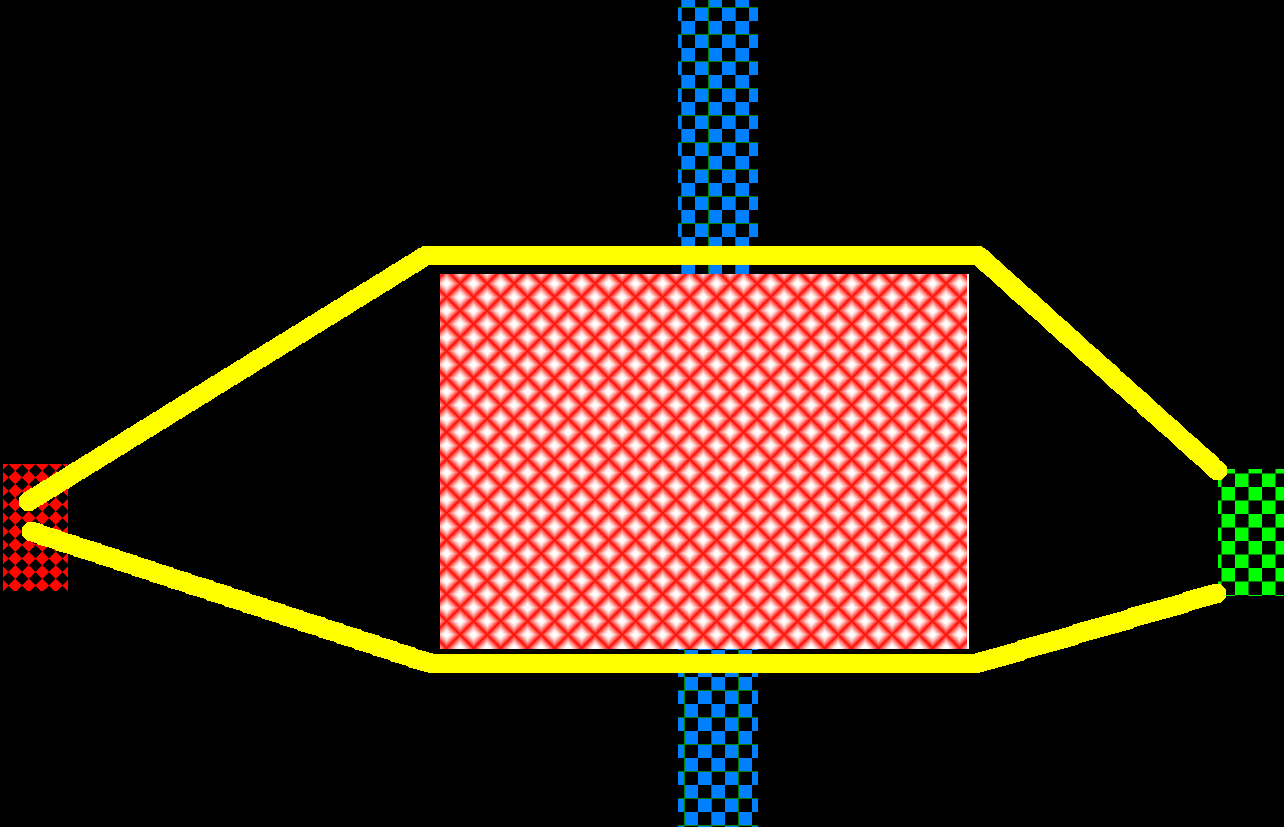} 
	\caption{Compared to figure \ref{fig:Example1010b} the black-blue areas mark intermediate destination areas. There are now two routes, one leading over each of the intermediate destination areas.}
	\label{fig:Example1010b}
\end{figure}

Imagine a walking geometry as shown in figure \ref{fig:Example1010}. In this case it is easy to make pedestrians pass the obstacle on either side by introducing on each side of the obstacle an intermediate destination area. This is shown in figure \ref{fig:Example1010b}. With the intermediate areas and the routes which include them, it is possible to route pedestrians locally into the direction of the shortest path, but still make a given fraction of pedestrians detour.

\begin{figure}[htbp]
  \center
	\includegraphics[width=0.612\textwidth]{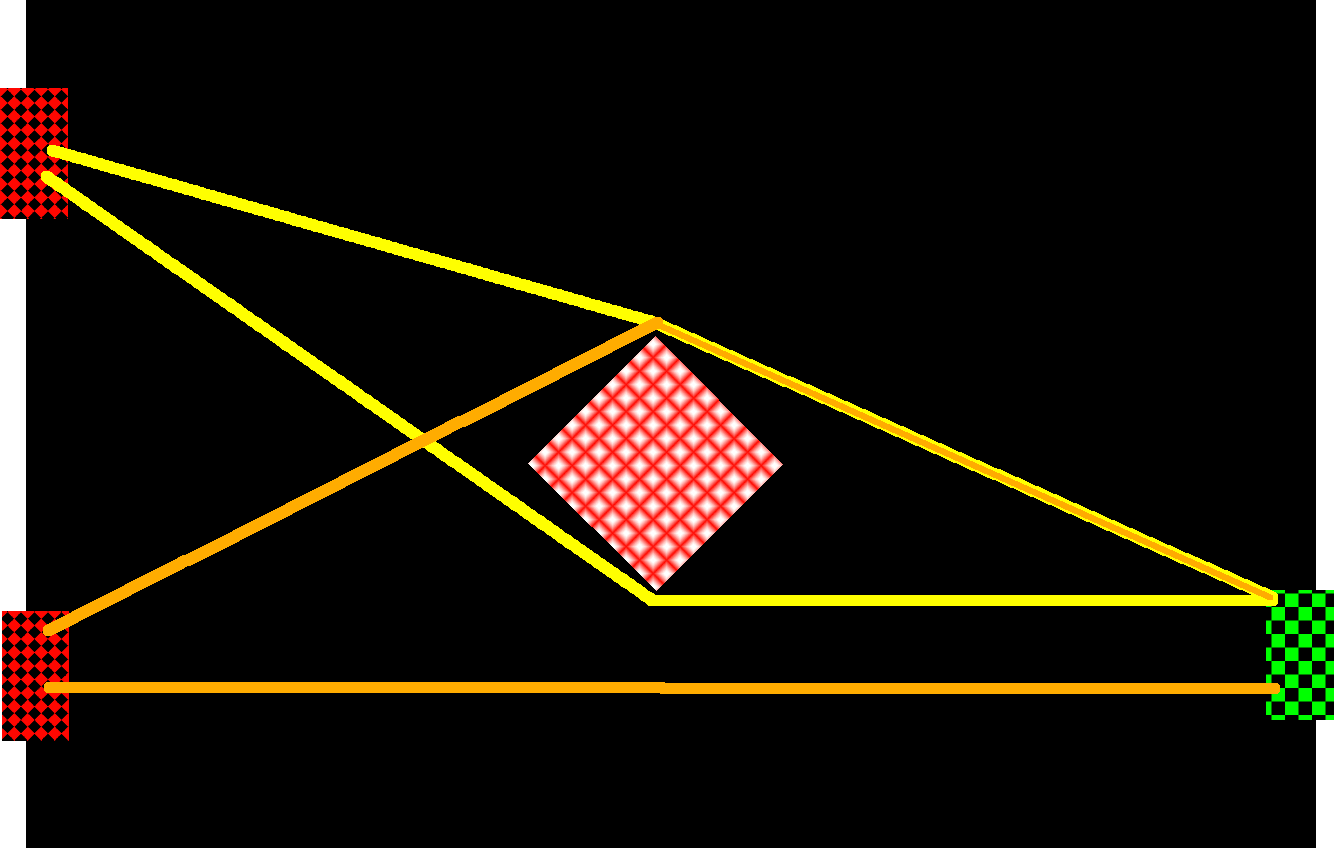} 
	\caption{Example with necessarily non-trivial geometry for the intermediate destination areas. Walking area (black), obstacle (diagonal red stripes on white ground), two origin areas to the left (red and black diagonal checkerboard), destination area to the right (green and black checkerboard), and shortest paths (yellow and orange). Compare figure \ref{fig:Example1009bcd}.}
	\label{fig:Example1009}
\end{figure}

In figure \ref{fig:Example1010b} the lower intermediate destination area was not necessary. Having a route directly leading from the origin to the destination area would have the same effect as long as pedestrians basically follow the direction of the spatially shortest path. In this case the lower intermediate destination area does not change the path of the pedestrians on that route compared to the case without any intermediate destination area. This is not in general the case. In general it is difficult to shape the intermediate destinations such that the path on the in principal shortest route is not distorted compared to the case without intermediate destination. Figure \ref{fig:Example1009} shows such a case. It is concluded that the intermediate destination areas cannot be of trivial shape (rectangles) in general.

\begin{figure}[htbp]
  \center
	\includegraphics[width=0.3\textwidth]{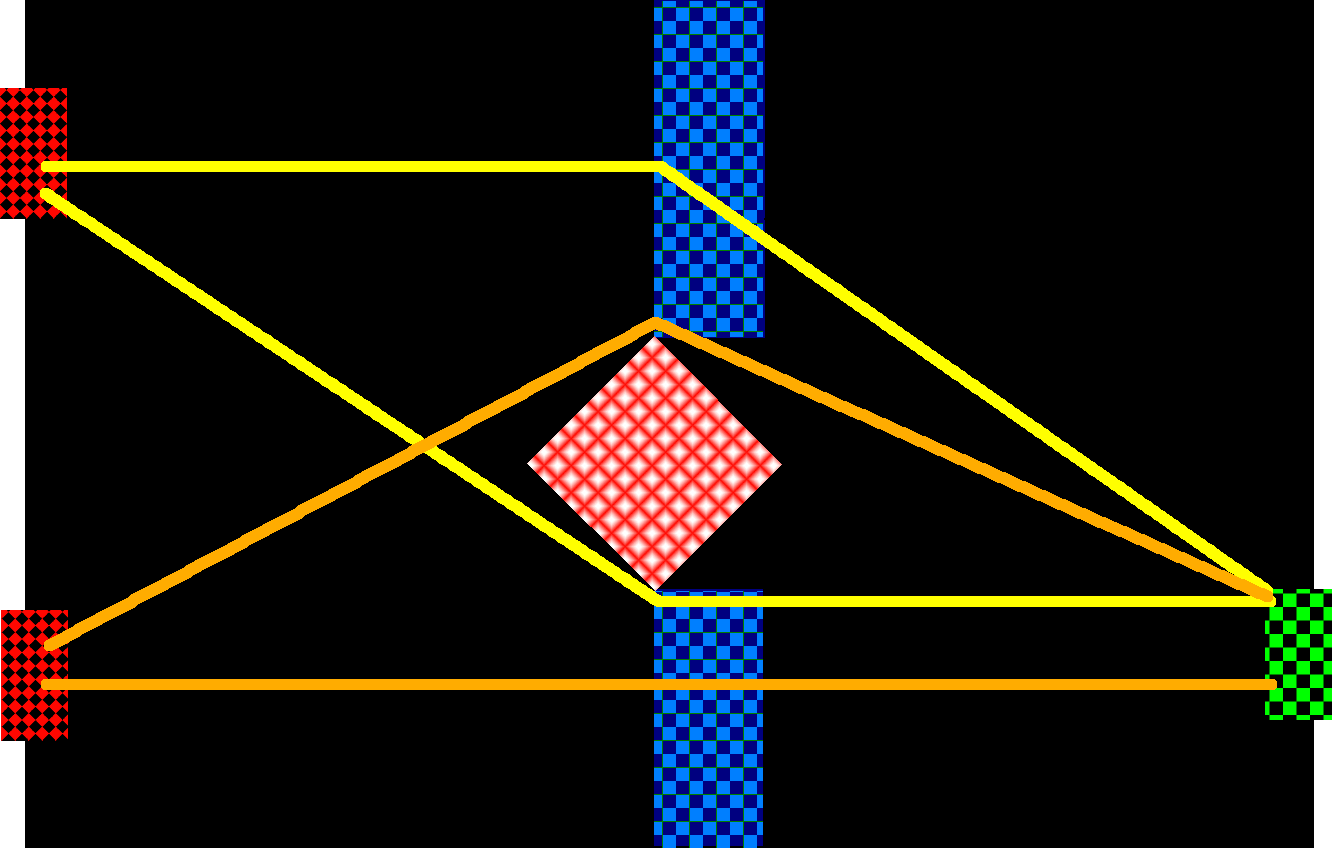} \hspace{6pt}
	\includegraphics[width=0.3\textwidth]{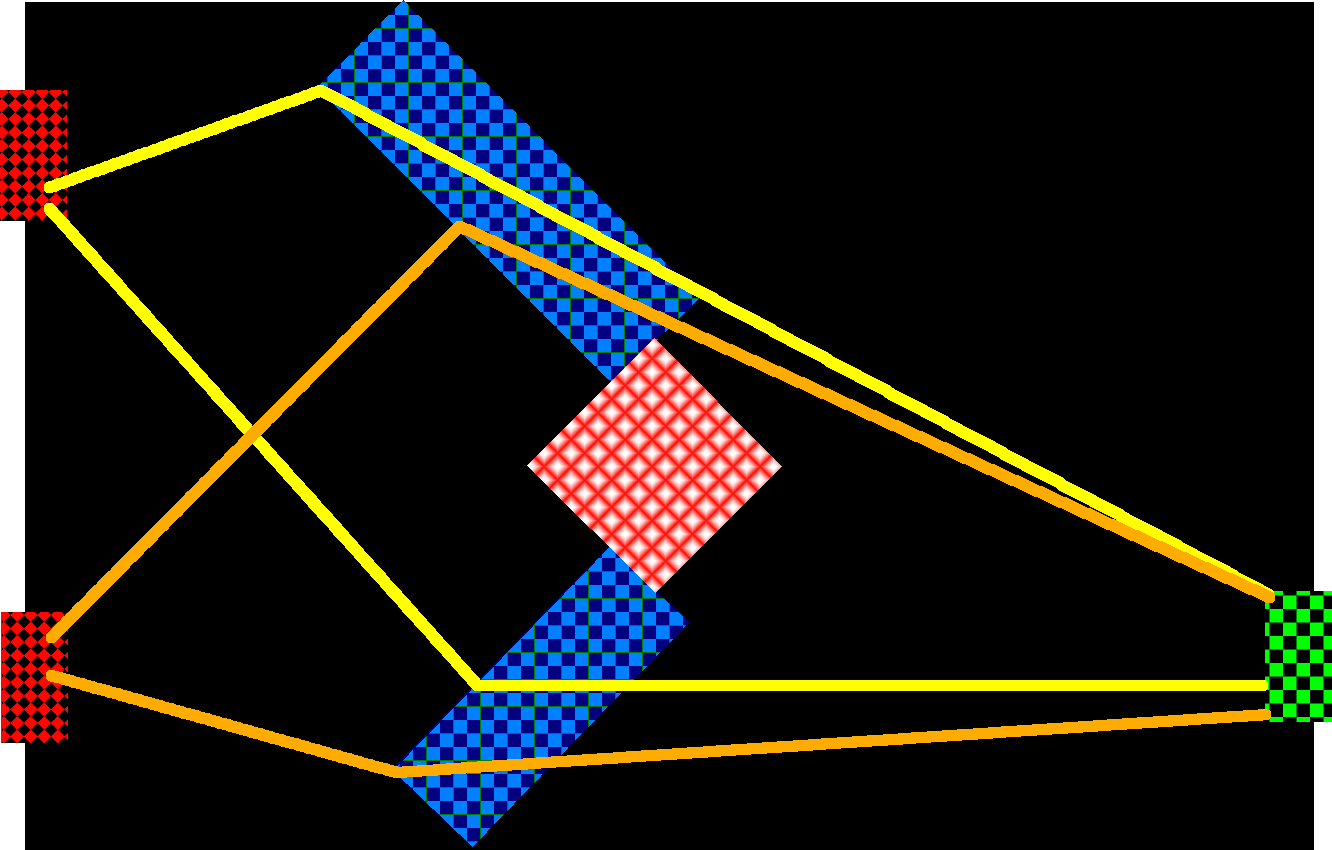} \hspace{6pt}
	\includegraphics[width=0.3\textwidth]{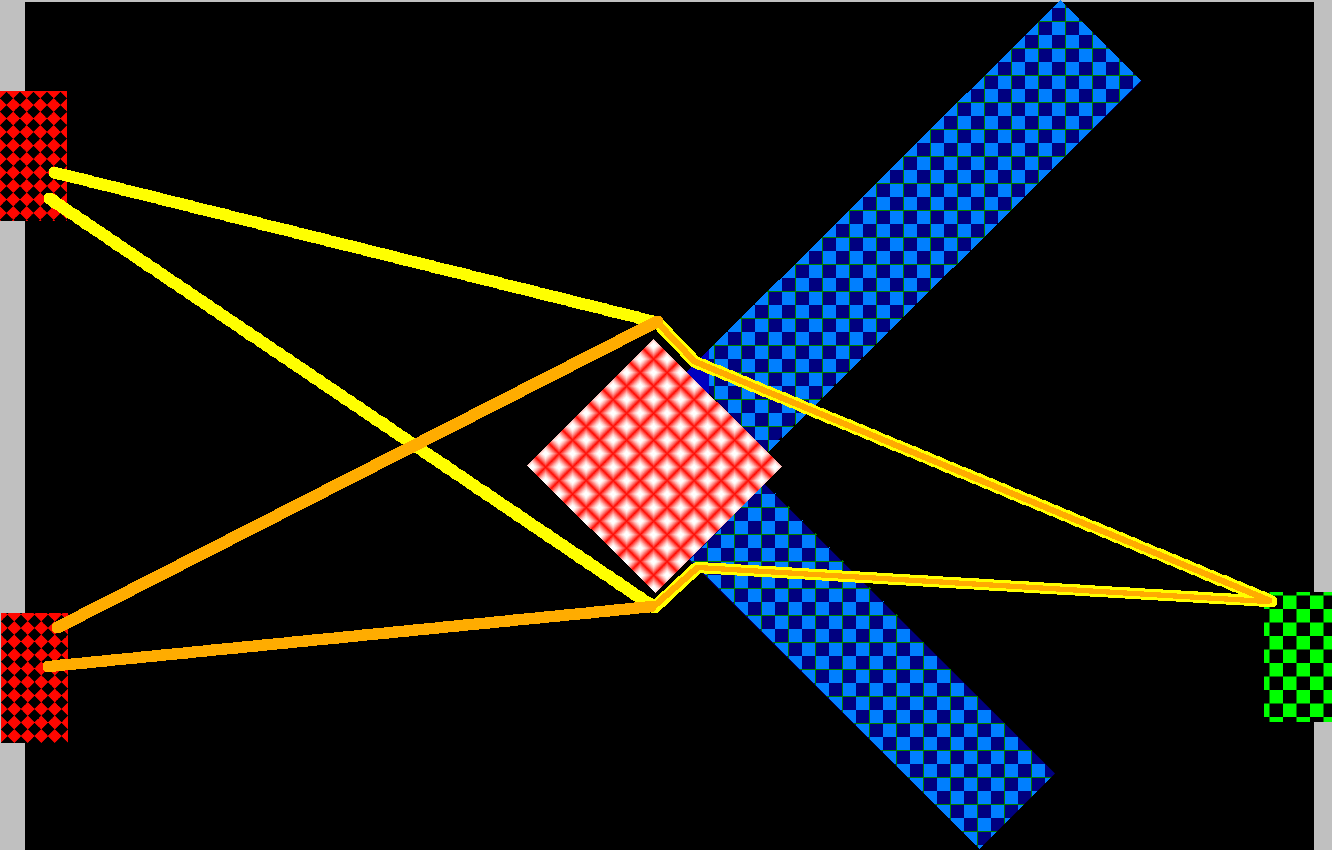} 
	\caption{For the example of figure \ref{fig:Example1009} simple rectangular areas are used as intermediate destination areas (light and dark blue checkerboard pattern). With none of the three variants the shortest paths as shown in figure \ref{fig:Example1009} are reproduced.}
	\label{fig:Example1009bcd}
\end{figure}

\begin{figure}[htbp]
  \center
	\includegraphics[width=0.4\textwidth]{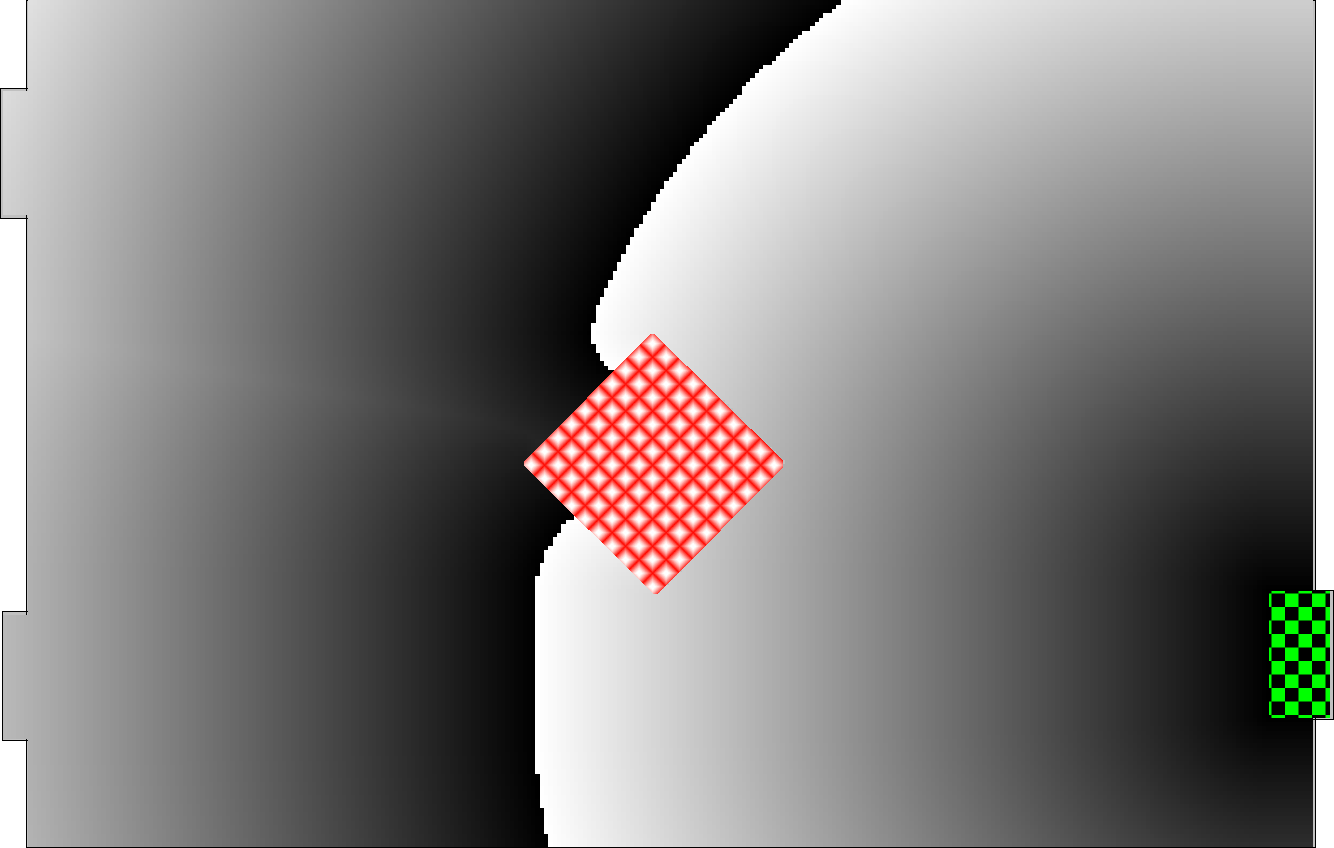} \hspace{6pt}
	\includegraphics[width=0.4\textwidth]{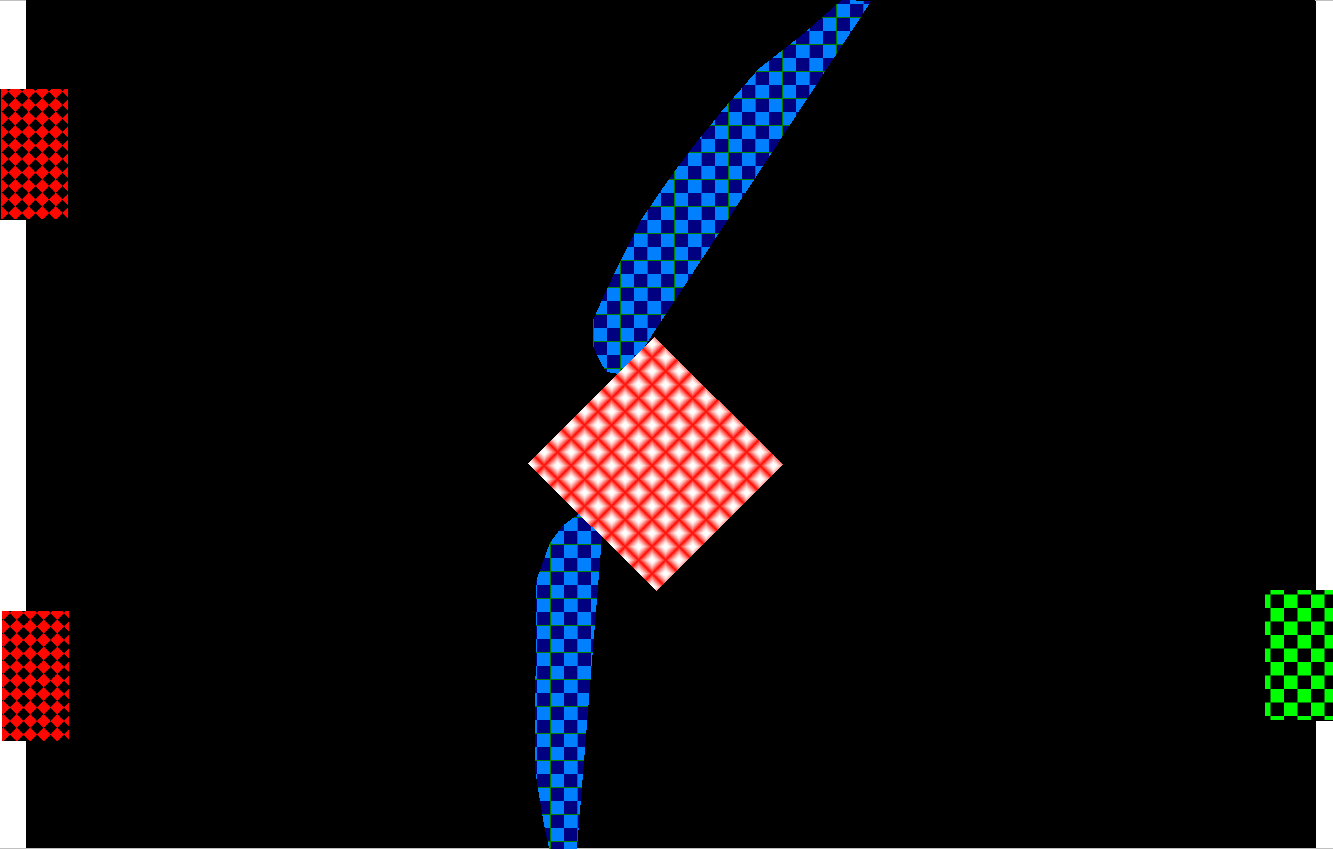}
	\caption{The left side shows the distance map of the destination area. Brighter pixels show larger distance. However, about at half the maximum distance the brightness is set back to zero to better display one distinct equi-distance line. The right side shows the geometry with two intermediate destination areas added which are shaped along that particular equi-distance line with their upstream (left) side. The exact shape of the downstream (right) side is not relevant. With these intermediate destination areas the shortest paths of figure \ref{fig:Example1009} are reproduced necessarily and exactly up to the limited precision in the numerical definition of the intermediate destination areas (smooth curves need to be approximated with polygons).}
	\label{fig:Example1009e}
\end{figure}

The basic idea is now that the intermediate destination areas need to be shaped along the equi-distance (equi-potential) lines of the destination area distance map (static potential). This is shown in figure \ref{fig:Example1009e}. For if an intermediate destination area is shaped along the equi-distance lines of the next downstream (intermediate) destination area the equi-distance it produces itself are -- in some surrounding of the intermediate destination -- in line with the ones of the next downstream (intermediate) destination. This implies that the gradients of both distance fields locally are identical. Thus the preferred walking direction remains unchanged no matter which of the two (intermediate) destinations a pedestrian is heading for.

\section{Method Formulation}
\begin{figure}[htbp]
  \center
	\includegraphics[width=0.612\textwidth]{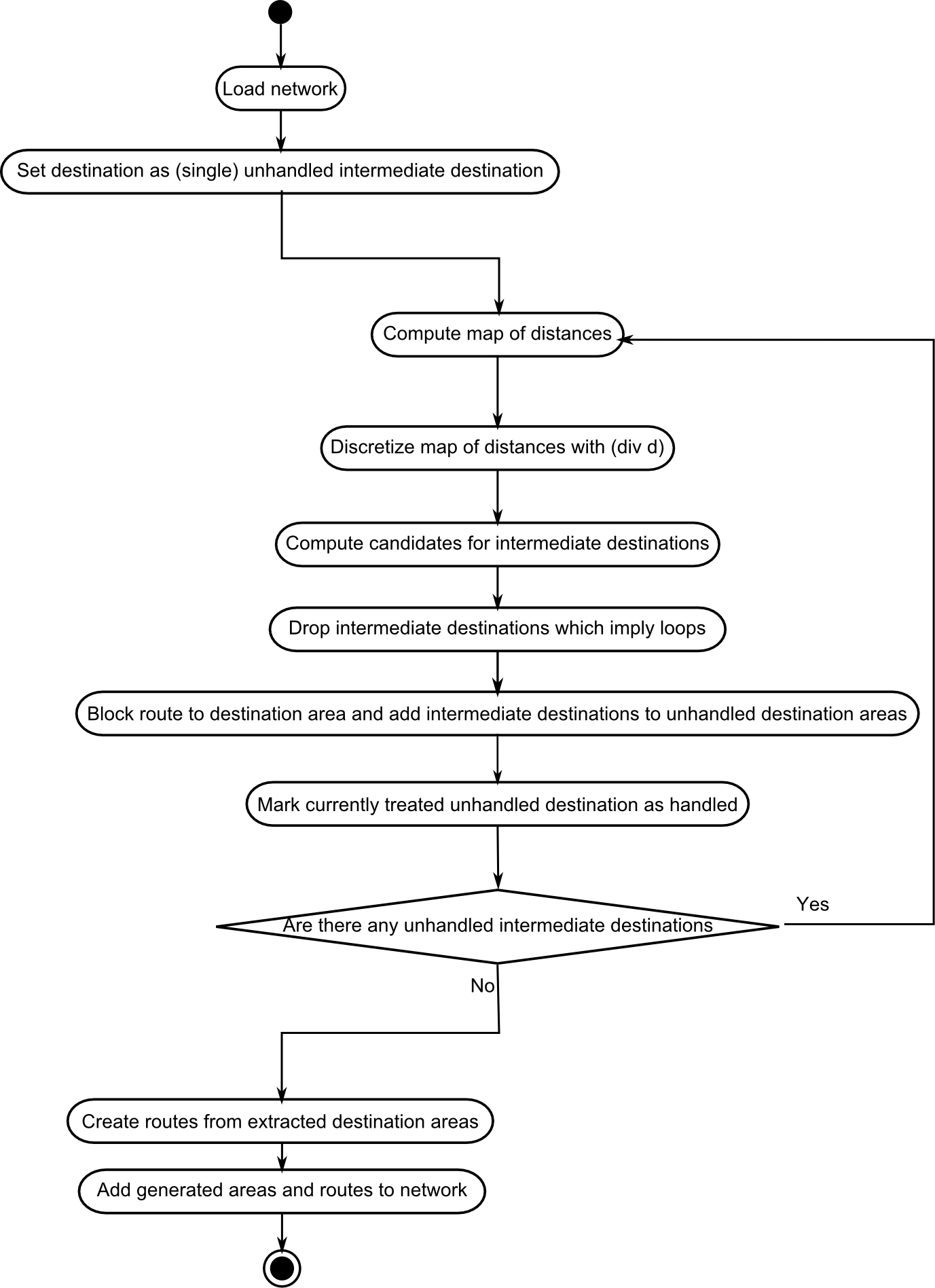} 
	\caption{Overview of the algorithm. The intended part is applied for each unhandled intermediate destination area (all related to one downstream intermediate destination area). The loop which is indicated marks the recursion (proceeding intermediate destination by intermediate destination from final destination owards origin).}
	\label{fig:flowchart}
\end{figure}

This section introduces a method for usage in simulations of pedestrian dynamics that computes for a given origin and a given destination area the shape of intermediate destination areas. as well as the routes which these are a part of. The method includes a parameter which determines the approximate diameter above which an obstacle or a hole in the walking area creates a routing alternative. Figure \ref{fig:flowchart} shows as an overview the steps which are explained in the following subsections.

\subsection{Basics}
\begin{figure}[htbp]
  \center
	\includegraphics[width=0.305\textwidth]{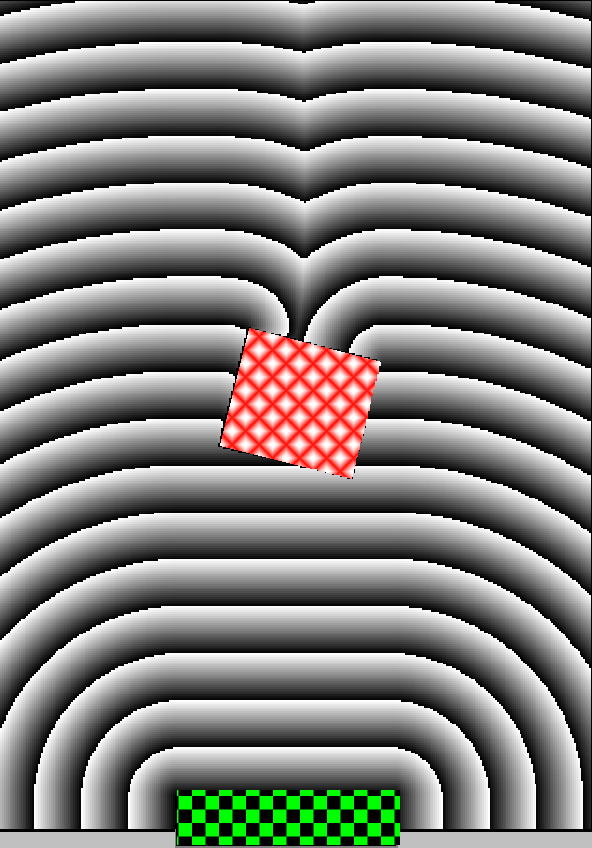} 
	\includegraphics[width=0.305\textwidth]{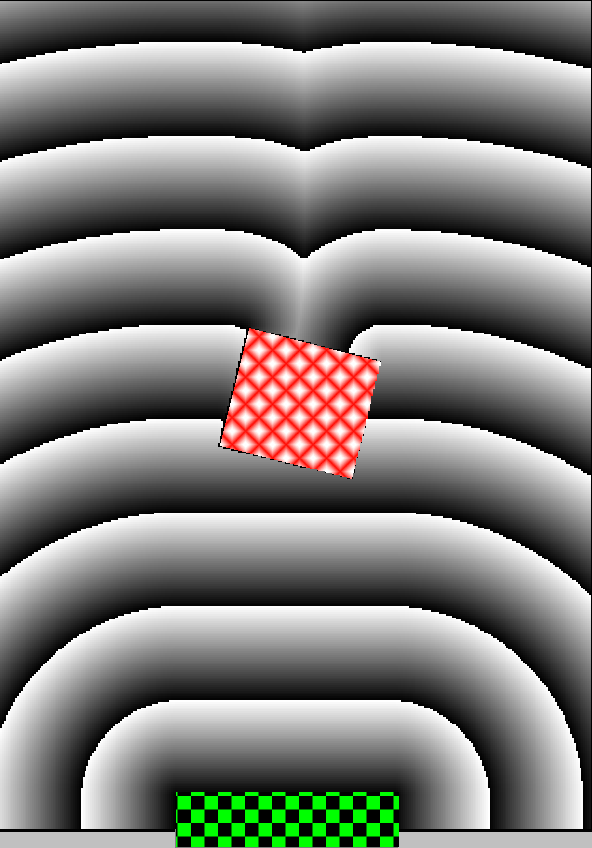} 		
	\includegraphics[width=0.305\textwidth]{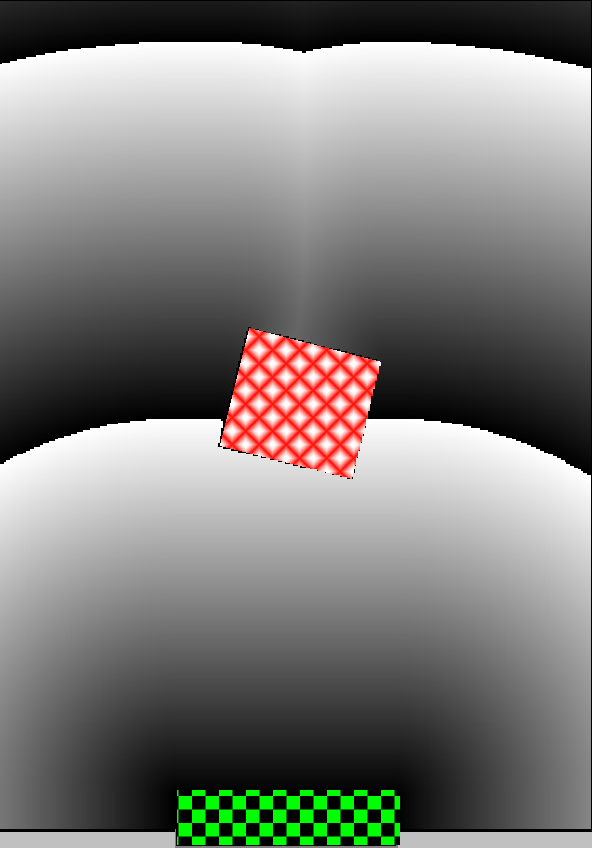} 	
	\caption{One potential displayed with three different modulo values $d$. The ratio of the modulo values of the middle and the left display is 2, the one of the right vs. the left one is 8. The destination is colored as black and green checkerboard, diagonal red lines on white ground depict an obstacle.}
	\label{fig:potentialmodulos}
\end{figure}

The first step is to compute a distance map for the destination (respectively for all destinations; this option is implied in the following) for which alternative routes ought to be computed. Next the modulo display of this potential as it is shown on the right side of Figure \ref{fig:potential} is computed. This is not only a useful display to see the properties of the distance map clearer, but its structure also plays a crucial role in finding the intermediate destination areas. Figure \ref{fig:potentialmodulos} shows the same potential for various modulo values $d$.

\begin{figure}[htbp]
  \center
	\includegraphics[width=0.305\textwidth]{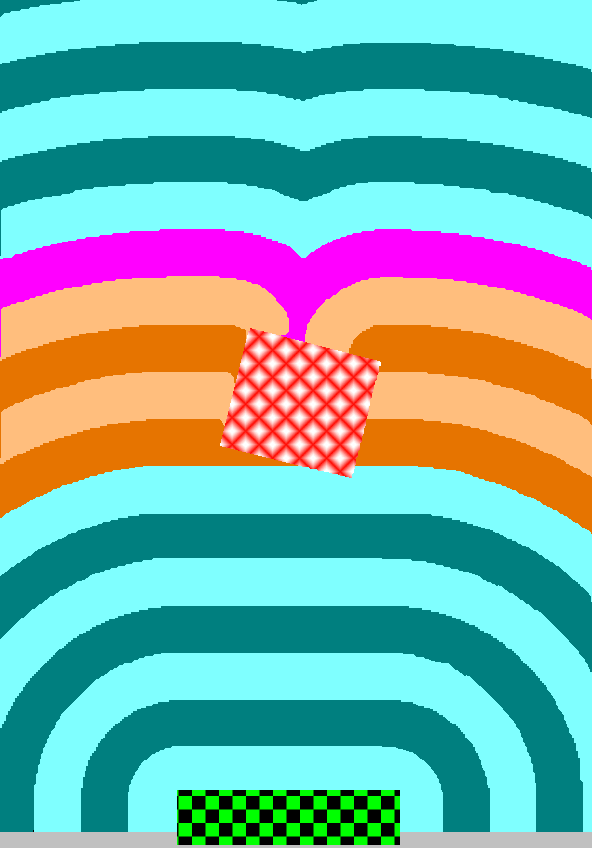} 
	\includegraphics[width=0.305\textwidth]{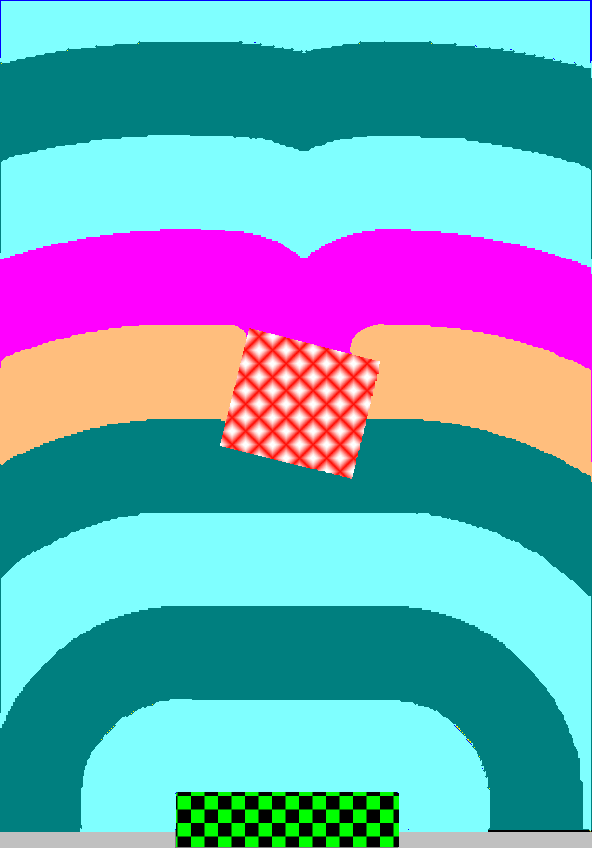} 		
	\includegraphics[width=0.305\textwidth]{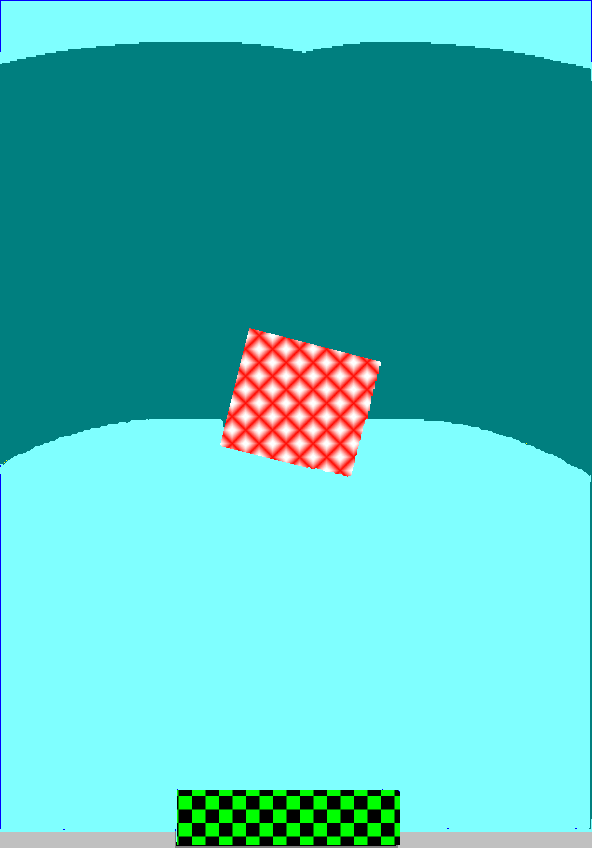} 	
	\caption{The destination is shown as green and black checkerboard, an obstacle is colored with diagonal red lines on white ground. Light and dark cyan mark regions which are simply connected and which lie within a range of distances to destination and which have exactly one simply connected area as direct neighbor which is closer to destination. Light and dark orange mark regions where there is more than one (here and usually: two) unconnected (split by an obstacle) regions which are within a range of distances to destination. The magenta region is simply connected and has more than one (here: exactly two) directly neighboring areas which are closer to destination than the magenta region is itself.}
	\label{fig:potentialmodulosB}
\end{figure}

Figure \ref{fig:potentialmodulosB} shows a modification of Figure \ref{fig:potentialmodulos}. The gradual color change from black to white has been omitted. Instead one range from black to white is shown in one and the same color and the next range in a different one. In this way areas which are within some distance range from the destination are grouped.

The coloring of Figure \ref{fig:potentialmodulosB} (left and center figure) is the key for the further steps in the method. All regions which share the property of the magenta colored region have to be identified. These are regions, which are unique for their range of distances to destination, but which have two unconnected regions directly neighbored which are closer to the destination than they are. 

Relevant origin positions of a pedestrian in a simulation of the scenario of Figure \ref{fig:potentialmodulosB} are the light and dark orange and the magenta regions as well as the light and dark cyan regions which are further away from the destination than the magenta area. If a pedestrian's origin position is on a cyan area which is closer to the destination than the closest orange area no route alternatives are generated.

In almost all simulation models of pedestrian dynamics the base direction is set into the direction of the shortest path. This implies that it is determined by the origin position if a pedestrian passes the (red) obstacle to the right or to the left side. If we want to choose freely if a pedestrian passes the obstacle either to the left or the right side, the magenta region in Figure \ref{fig:potentialmodulosB} is the critical one. Once a pedestrian has passed it on the intended side of the obstacle, the shortest path paradigm will take the pedestrian on this very side to the destination; the pedestrian will not turn around anymore and pass the obstacle on the other side.

Therefore the two light orange regions immediately neighbored to the magenta region are made intermediate destination areas. Each of the two is assigned to one route, generating a route choice. Sending a pedestrian over one of the two routes ensures that he passes the obstacle on the intended side.

Obviously not only the shape of the additional intermediate destination areas depends on parameter $d$, but first and foremost if it exists at all. The right figure of Figure \ref{fig:potentialmodulosB} shows no additional intermediate destination areas because the value of parameter $d$ has been chosen too large. 

Parameter $d$ therefore has the role of a wave length: just as an object cannot be seen if it is smaller than the wave length of light which is shed on it. Just as a ship which is smaller than the wave length in the surrounding sea is just going up and down but not disturbing and reflecting the waves, an obstacle which is smaller than the value of parameter $d$ remains invisible in the sense that it has no effect (of creating an additional intermediate destination).

The choice of the value of parameter $d$ at first might appear to be arbitrary and its existence therefore as a major downside of the method. However, first, simulation models of pedestrian dynamics usually contain elements to deal with obstacles: pedestrians walking around them, keeping some distance, etc.. For small obstacles these elements can make pedestrians pass on either side even if the model of pedestrian dynamics has no elaborate route choice method included. Thus, for small obstacles no additional route choice method is required. Second, for the purpose of computational efficiency and limited computation resources, in large simulation projects it can be desirable to not consider every existing routing alternative, but only the major ones. Then one may set the value of parameter $d$ to larger values to maintain a feasible simulation efforts. 

At the very end of this contribution we show that there are geometries where it would be favorable to have a small value for parameter $d$. However, for the reason to be able to select the size of obstacles which create route alternatives and because we think that the approach with the areas created by integer division of distances is very illustrative we stick with it and will discuss later also how to fix the emerging problem.

The left side of Figure \ref{fig:potentialmodulosB} suggests that there is some arbitrariness in selecting the regions which are directly neighbored to the magenta region as intermediate destination areas. Why not two of the dark orange or the other two light orange regions? The answer can be found by considering pedestrians who have an origin position somewhere on the dark orange or light orange regions. It is desired to be able to also assign such pedestrians a route which leads over an intermediate destination area on the opposite side of the obstacle. If that intermediate destination area was moved toward the destination, some pedestrians would approach the intermediate destination from the back side. This implies an unrealistic trajectory and is therefore not desired. In fact because any intermediate destination area has a finite depth, there can be a small area where this occurs although the region closest to the magenta region has been made an intermediate destination area. This problematic region is the larger, the larger the value of parameter $d$ is. Realizing this it is important to realize as well that only the shape of the front side of an intermediate destination is relevant. The function of parameter $d$ is to determine existence of intermediate destination areas not their depth. The depth can be reduced to the point that it is still guaranteed that any pedestrian heading to the intermediate destination does not have to slow down to step on it for at least one time step. Nevertheless for the remainder of this contribution we will not modify the shape of intermediate destination areas in this way. The unmodified shape simply gives a clearer impression of the idea.

\subsection{Multiple Obstacles} \label{sec:MO}

In the previous subsection the basic idea of the method was demonstrated in an example with just one obstacle. One has to give some additional thoughts on the usual case with more than one obstacle.

\begin{figure}[htbp]
  \center
	\includegraphics[width=0.612\textwidth]{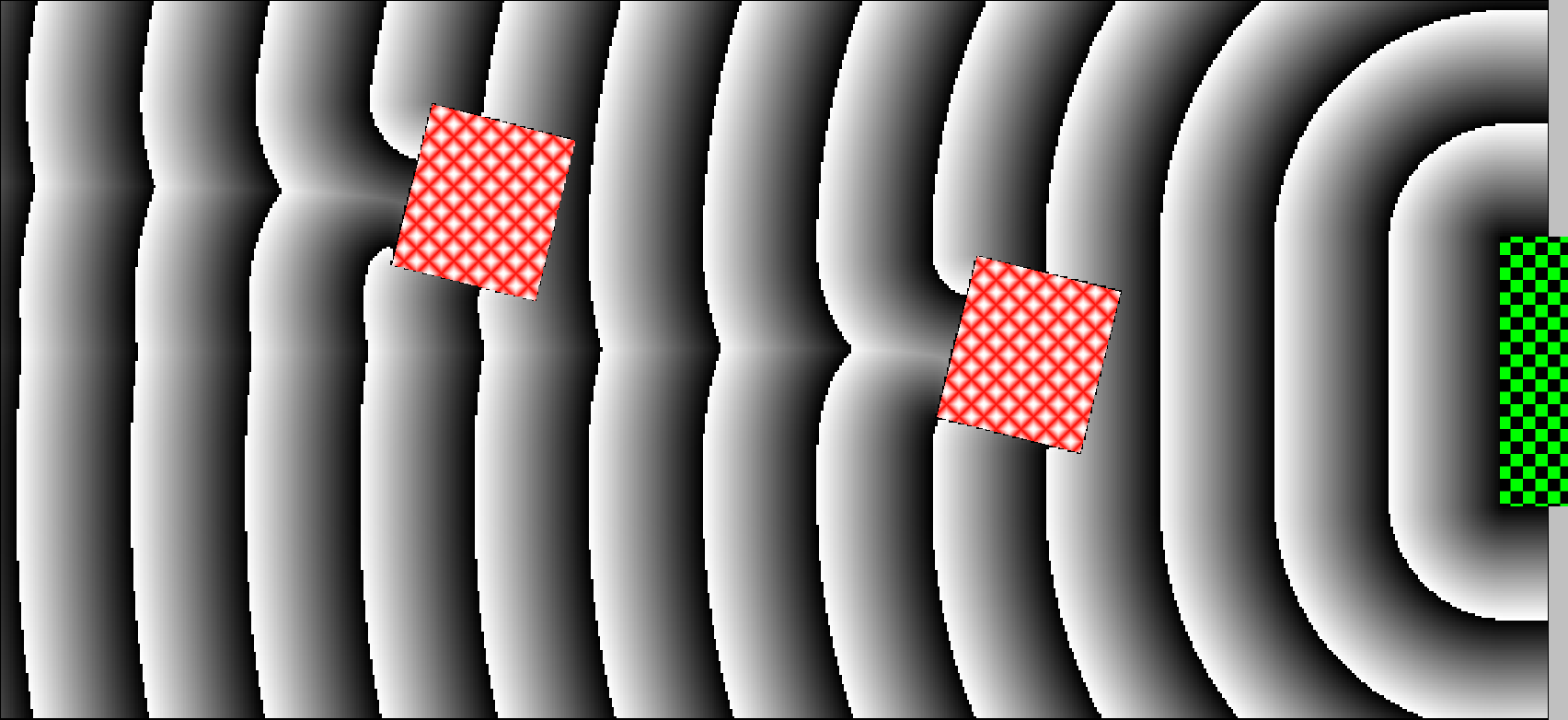} \\ \vspace{6pt}
	\includegraphics[width=0.612\textwidth]{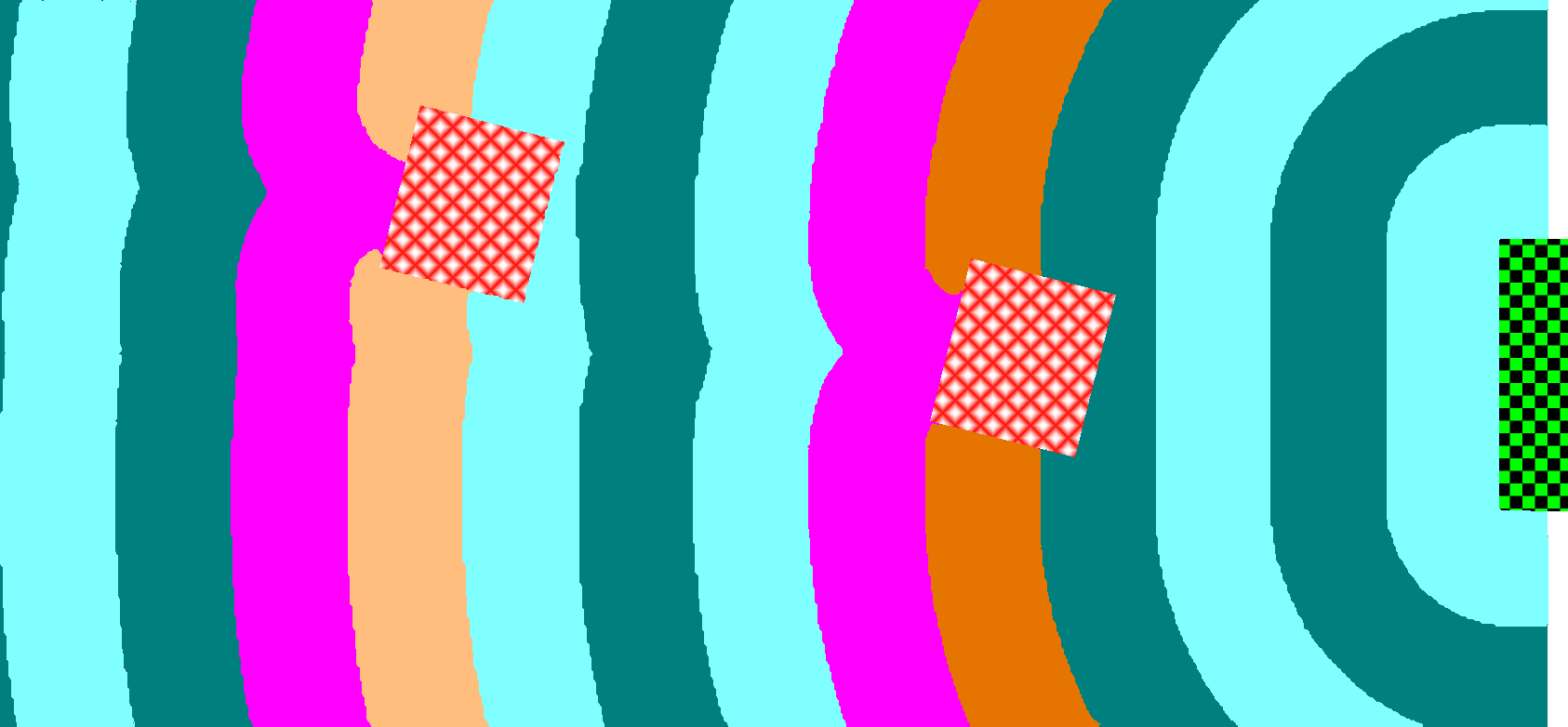} 		
	\caption{Compared to the previous example a second obstacle has been added.}
	\label{fig:extended1}
\end{figure}

\begin{figure}[htbp]
  \center
	\includegraphics[width=0.612\textwidth]{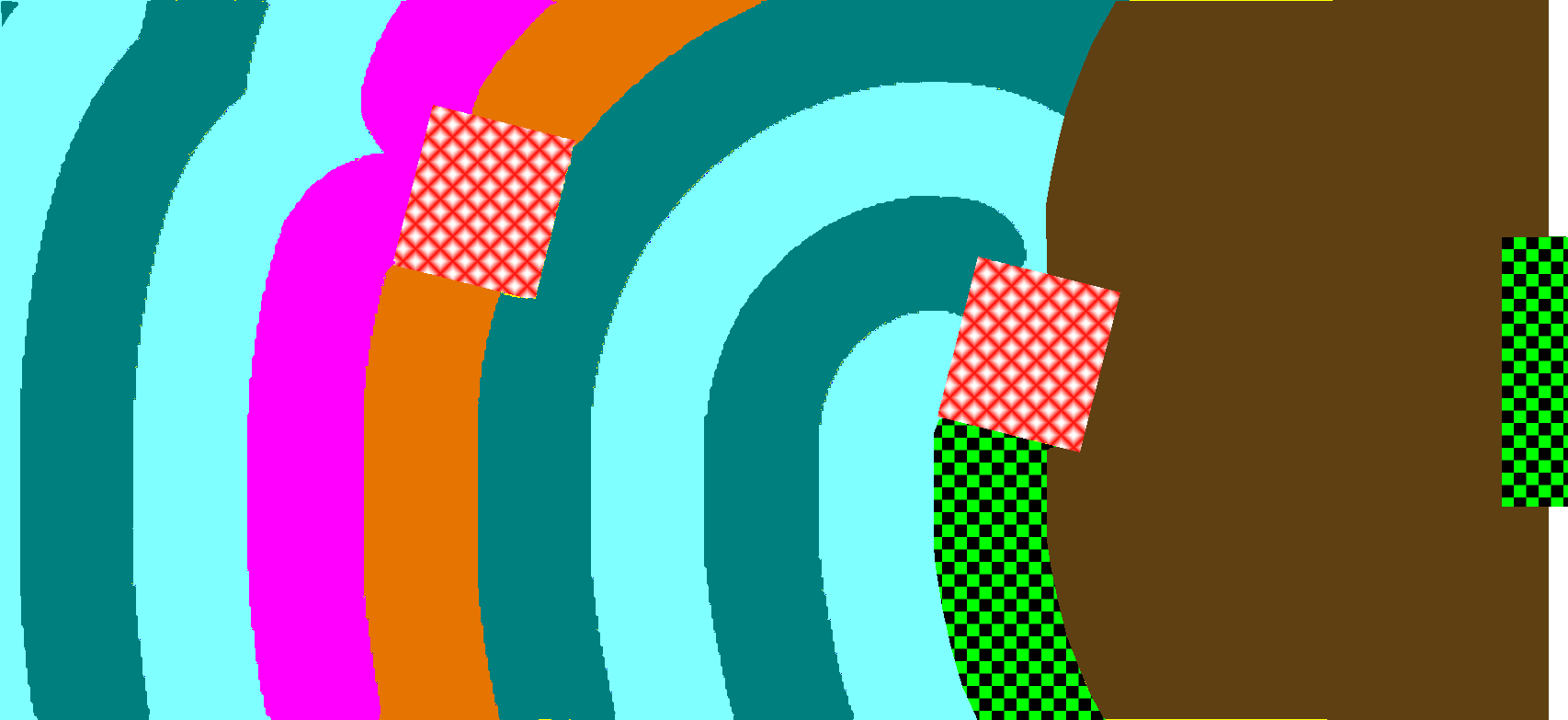} 		
	\caption{Computation of the upstream intermediate destination areas with respect to one of the downstream intermediate destination areas. The region where pedestrians' origin positions do not generate route alternatives is shown brown. It is a virtual obstacle into which in the further recursions of the algorithm not anymore a potential (resp. distance map) will spread.}
	\label{fig:extended2}
\end{figure}

\begin{figure}[htbp]
  \center
	\includegraphics[width=\textwidth]{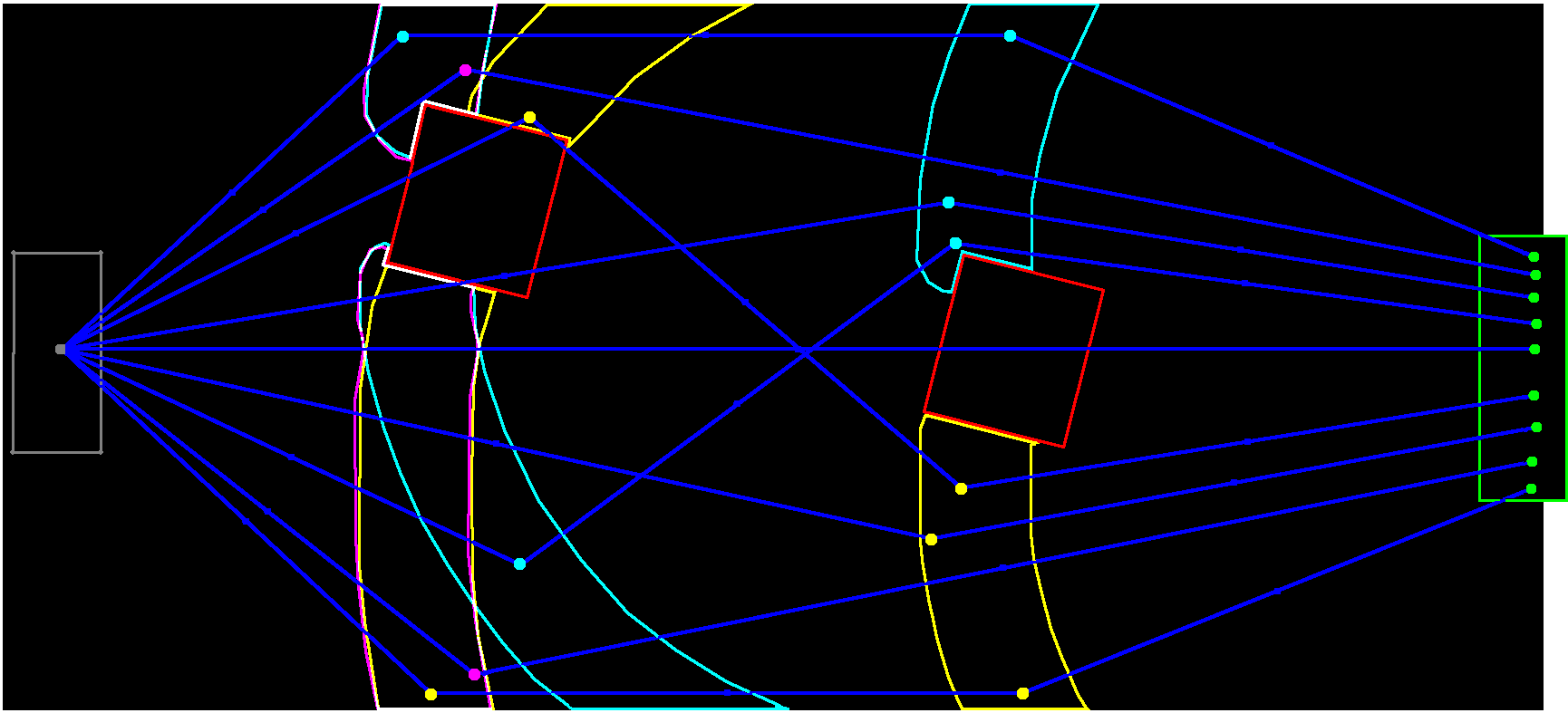}		
	\caption{Shown here are the edge of the destination area (green), the edges of the obstacles (red), the edge of some origin area (gray), the edges of the downstream intermediate destination areas (cyan and yellow) and the upstream intermediate destination areas (cyan, yellow, magenta). Upstream and downstream intermediate destination areas which follow in a sequence on one route are shown in the same color (cyan or yellow). Furthermore this figure shows the routes in blue. The origin area is marked with a gray dot, the destination area with green dots. All intermediate destination areas are marked with the color that corresponds to the color of the edge of the area where they are attached. The dots only mark the area which is an intermediate destination area. Their position on the area has no impact on the movement of pedestrians in the simulation. It is interesting to note that the magenta intermediate destinations share part of their edge with the cyan and another part with the yellow intermediate destinations (this makes the common edges partly appear white). This is not a coincidence, but a typical consequence of the algorithm.}
	\label{fig:extended3}
\end{figure}

If there is more than one obstacle in a scenario the basic idea presented above has to be applied recursively. This is shown in Figures \ref{fig:extended1} to \ref{fig:extended3}. Figure \ref{fig:extended1} shows the initial computation step for and with the potential. The dark orange regions are identified as intermediate destinations to pass the downstream obstacle, the light orange regions have the same role for the upstream obstacle. However, the shape of the upstream regions is computed with respect to the destination. If we want to be able to send pedestrians without artifacts in their trajectories on routes which lead from an upstream intermediate destination area to that downstream intermediate area which is more remote respectively, we have to calculate the shape of the upstream intermediate destination areas with respect to the downstream intermediate destination areas. For one of the two downstream intermediate destination areas this is demonstrated in Figure \ref{fig:extended2}

The brown region in Figure \ref{fig:extended2} -- the region which is closer to the destination than the downstream intermediate destination area -- marks the region where pedestrians' origin positions do not generate routing alternatives. Also when the map of distances -- the potential -- is computed beginning at the downstream intermediate destination area, the brown region must not be intruded. For the computation of routing alternatives -- not later in the simulation -- it is therefore treated as if it was an obstacle. If the brown region was intruded during the computation of the potential then short cut paths could result which would lead to unrealistic movement and thus spoil the intended effect of the method. 
Figure \ref{fig:extended3} finally shows all resulting intermediate destination areas and resulting routes.

In Figure \ref{fig:extended3} one can see that the original route without intermediate destinations is preserved -- it is the route which ends at the central of the nine destination dots which are depicted in green. Of the nine routes the one to the first green destination point from bottom, second, third, and fifth from bottom basically are identical as they either have no intermediate destination point or if they have then they are along the globally shortest path. 

The third, fifth, and seventh route from bottom (those with only one yellow marked intermediate destination, no intermediate destination, and only one cyan marked intermediate destination) can be identified easily within the algorithm as ones which multiply existing routes. As within the first iteration step on the search for upstream intermediate destinations four results are found (yellow and cyan downstream as well as the two magenta ones) one can erase the route without intermediate destination. As within the second iteration step starting at the downstream yellow (cyan) area two more intermediate destinations are found, one can erase the route with just one yellow (cyan) intermediate destination. Put in a different way: said three routes only result if one considers an intermediate destination area which was dealt with in the n-th recursion step also in the n+1-th recursion step, even if in the n-th step one or more upstream intermediate destinations were found.

What remains are the two magenta intermediate destinations. These cannot so easily be identified as routes which only double the effect of other routes as they could be required if route choice was nested (see next subsection).

For these remaining double routes the question is if they imply a problem or an inefficiency in an assignment method? It is surely not a serious problem with respect to assignment methods. Most assignment methods can deal with multiple identical routes. In realistic networks there are always routes which overlap. Assignment methods have to cope with this. Identical routes are simply routes which overlap by 100\%. 

Generally the recursive search for further upstream intermediate destination areas comes to an end when the potential (the distance map) covers the entire walking area\footnote{In principle one can already stop when the starting area (or all starting areas) is covered. For this, however, one has to know for sure when executing the algorithm where the starting area is. If this is not known or if it can change one has to consider the whole walking area.}, but no further area with properties as the magenta areas in the example is found.

The next two subsections consider further termination considerations for the recursion.

\subsection{Nested route choices}

Figure \ref{fig:extended3} raises the question, if the intermediate destination areas whose edges are colored magenta could be dropped together with their routes. Obviously the two routes do not actually add an alternative path choice. So is it sufficient to always just consider that routing alternative which is currently closest to destination (in Figure \ref{fig:extended1} marked by the dark orange areas in the lower figure) for further process and omit any more remote one (the light orange areas in Figure \ref{fig:extended1})?

\begin{figure}[htbp]
  \center
	\includegraphics[width=0.612\textwidth]{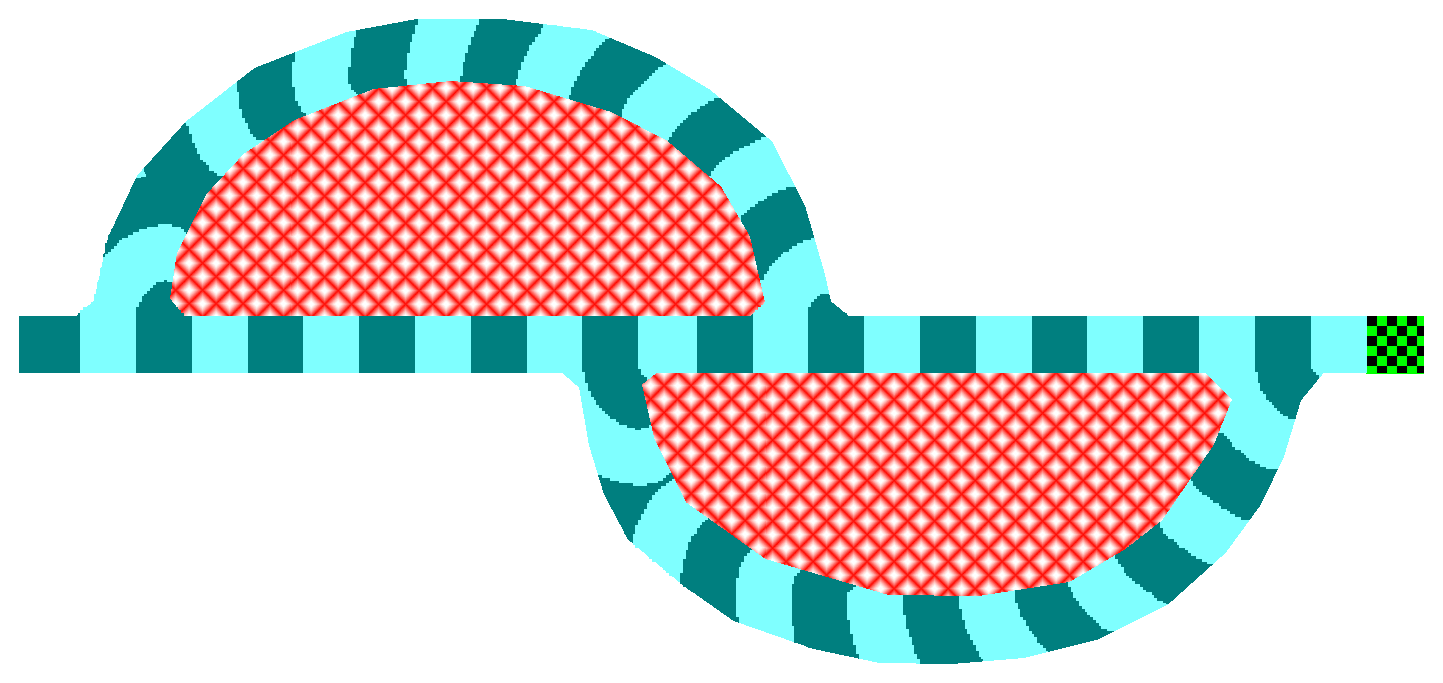} \\ \vspace{6pt}
	\includegraphics[width=0.612\textwidth]{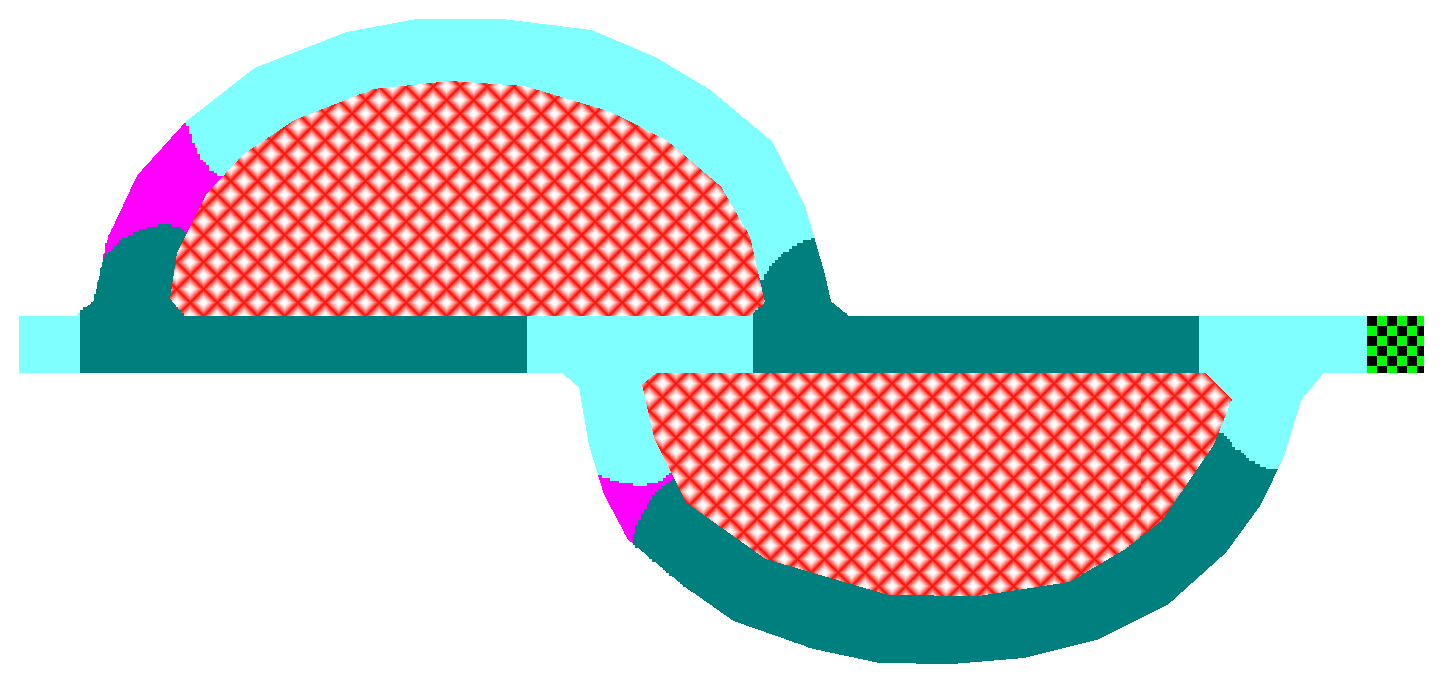} \\ \vspace{6pt}
	\includegraphics[width=0.612\textwidth]{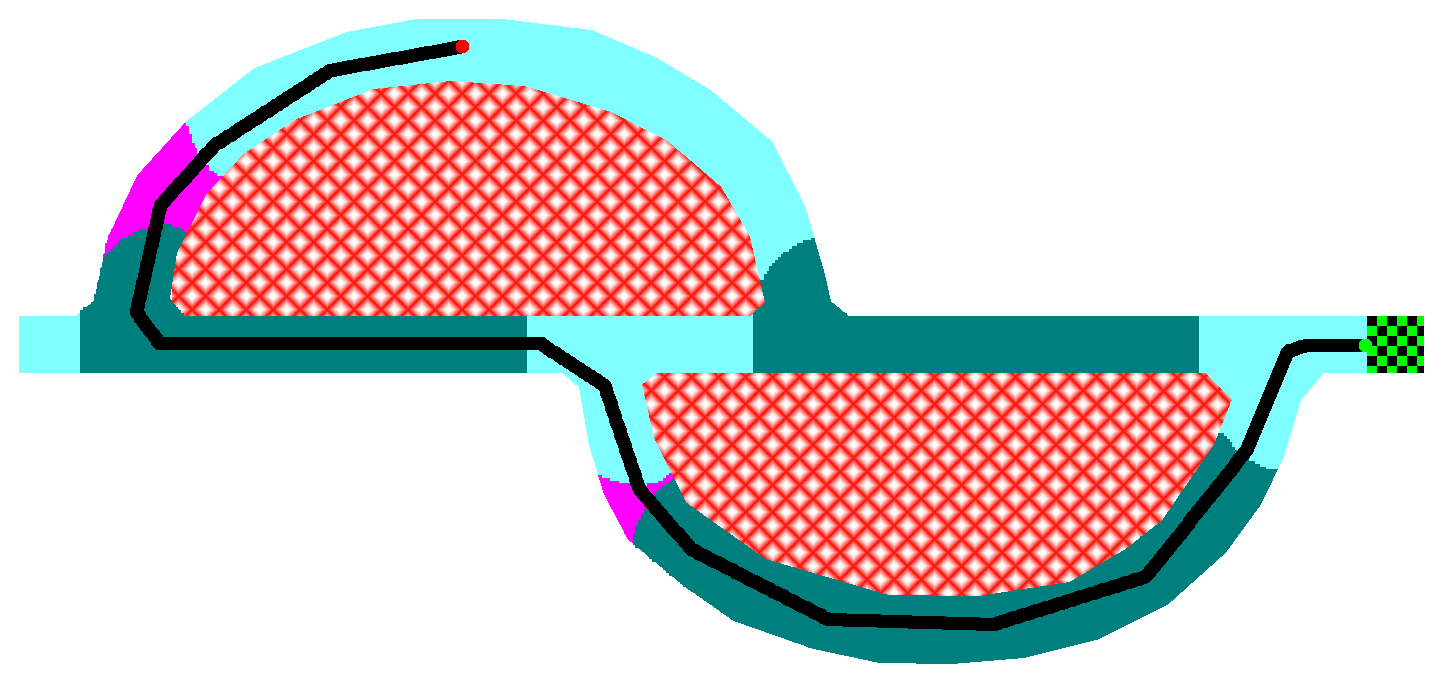} \\ \vspace{6pt}
	\includegraphics[width=0.612\textwidth]{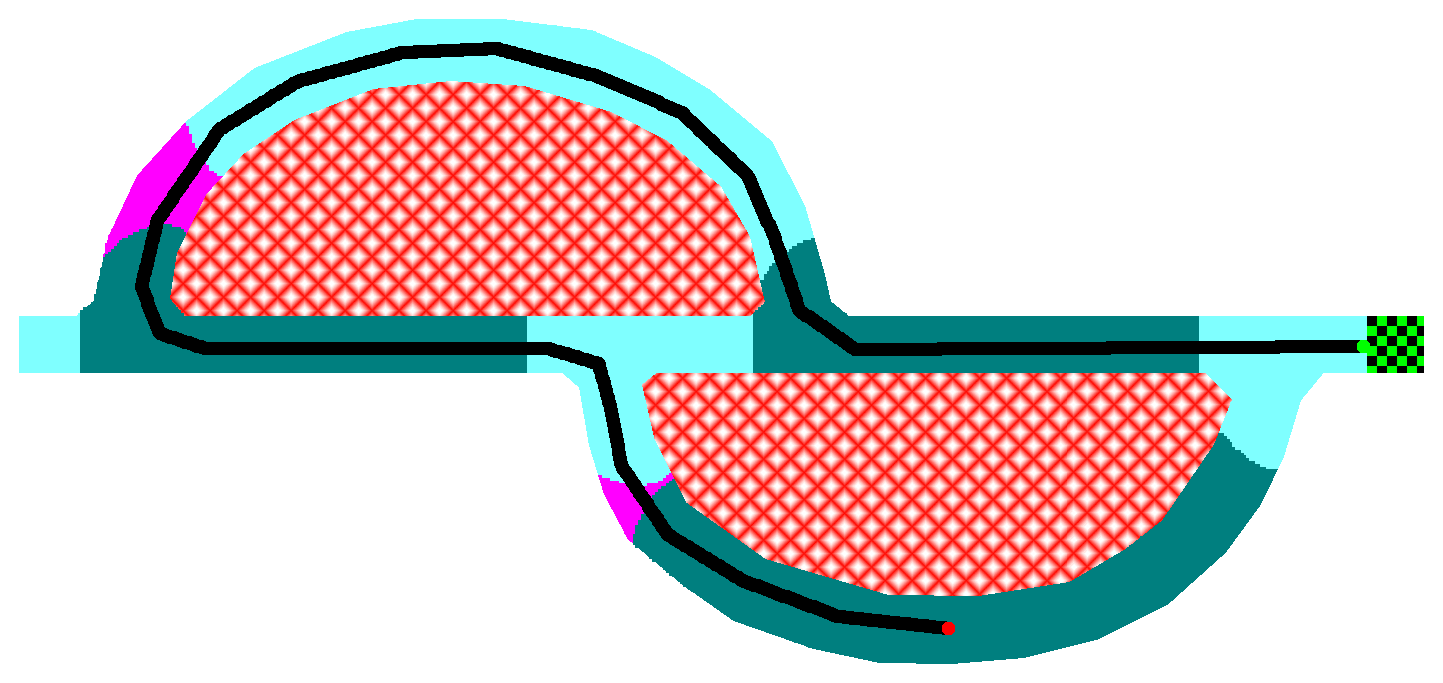}
	\caption{In this example the route choices are nested. The destination is on the right side. The first computation of the potential (first figure) brings up two route alternatives, respectively two local maxima of distance to destination (as it was the case in the example of the previous subsection). These are manifest by the existence of the two regions marked with magenta in the second figure. The third and fourth figure show two different possible trajectories (black; from red to green) of pedestrians where the two magenta regions are passed in different sequence.}
	\label{fig:Braess1}
\end{figure}

The answer is``no'' and the reason for this is that there are cases where route alternatives are nested; or in other words: it cannot necessarily be decided what is ``upstream'' and what ``downstream''. Figure \ref{fig:Braess1} demonstrates this with an example. Note that this example is not much different from the one in the previous subsection. The topological genus is the same. One can transfer one example into the other one by stretching and shrinking the geometry. 

It can -- of course -- always be decided which of two points is spatially closer to a destination. However, in terms of travel time this is different. One can easily imagine in the third and fourth figure of Figure \ref{fig:Braess1} where there has to be a jamming crowd of pedestrians to make the indicated path the quickest to destination although it implies and includes a spatial detour.

\subsection{Loops}

In reality people can walk loops. However, in planning tools -- be it for vehicular or pedestrian traffic -- one wants to exclude the emergence of loops from the algorithms unless the planner takes explicit measures to force loops occurring with a specified frequency. The method proposed so far produces sequences of intermediate destinations which lead to pedestrians walking in loops. This and how to avoid this will be discussed in this subsection.

The intermediate destination areas which are computed in the examples of this work share some properties: one could describe them as slightly distorted rectangles. Each rectangle has an upstream edge which is mainly adjacent to pedestrian walking space and in some cases partly to obstacles. The opposing side is the downstream side, also neighbored to walking spaces and maybe partially to obstacles. The other two mutually opposing edges -- and that is the relevant point here -- are neighbored to obstacles or more general to spaces where pedestrians cannot walk. These common properties are not accidental in the examples of this contribution. Only when the intermediate destination areas are clamped between two obstacles they clearly separate the local environment into upstream and downstream. In other words: only when two opposing sides ``touch'' obstacles the other two sides can take the role of an upstream and a downstream side, only than can the area implicitly define a direction. 

\begin{figure}[htbp]
  \center
	\includegraphics[width=0.4\textwidth]{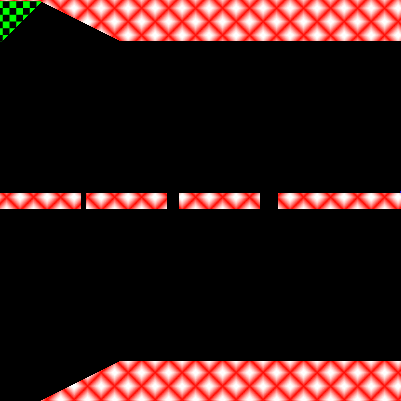} \hspace{6pt}
  \includegraphics[width=0.4\textwidth]{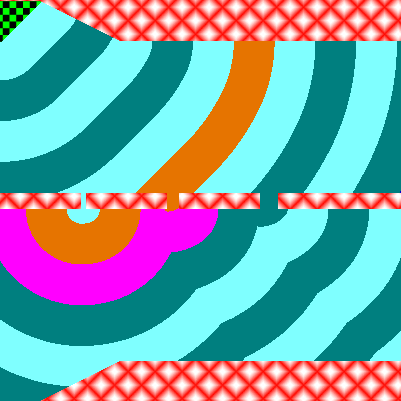} \\ \vspace{6pt}
  \includegraphics[width=0.4\textwidth]{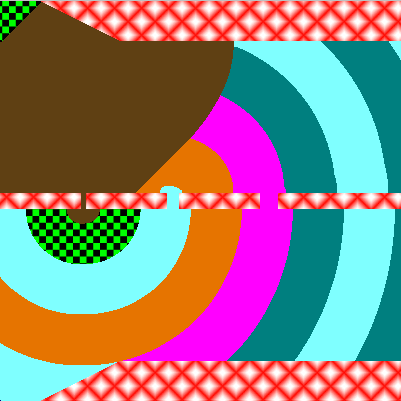} \hspace{6pt}
  \includegraphics[width=0.4\textwidth]{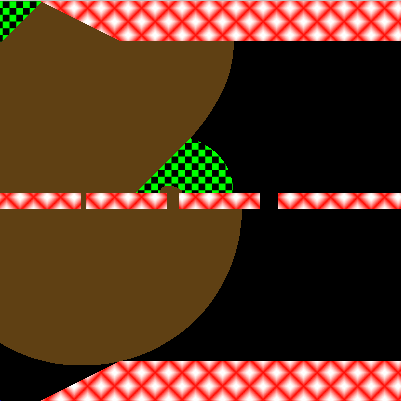} 
	\caption{The upper left figure shows an example geometry with the destination in the upper left corner. The origin could for example be located in the lower left or right corner. Upper right figure: in the first step the distances are computed for the destination. A route alternative emerges and with it two areas -- marked orange -- which in later routes would become intermediate destination areas. Lower left figure: computing the distance maps for one of the two intermediate destination areas brings up another route alternative. Lower right figure: one of the two intermediate destination areas that have emerged in the previous step has only one common edge with a real obstacle. The opposite edge is (partially) immediately adjacent to downstream area -- marked brown. This is considered to be the equivalent of a route including a loop which is prohibited by definition. Therefore this branch is not followed further and no route is generated. However, the other intermediate destination area shown in the lower left figure may well result in a route.}
	\label{fig:loop1}
\end{figure}

This raises the question how to deal with intermediate destinations where one or even both ``obstacle edges'' do not touch real obstacles, but a virtual obstacle -- an area as the one marked brown in Figure \ref{fig:extended2} -- emerging from the algorithm itself. Figure \ref{fig:loop1} shows that and how this can actually happen.

If such an intermediate destination area emerges from the algorithm it can and needs to be ignored together with the route which is linked with it. The algorithm recursion can terminate at this point. The reason for this is that such a construct can be interpreted as a loop in the route. The virtual obstacle in fact is a more downstream region. An intermediate destination area which is immediately neighbored to such a region implies that a pedestrian who is located on the intermediate destination area immediately next to the virtual obstacle with one small side step could skip a part of his route, respectively if he or she still would carry it out would walk a loop. Any pedestrian who is located somewhere else on the intermediate destination area could move on the intermediate destination area toward the virtual obstacle area. This should always be possible as intermediate destination areas are constructed to guide around bottlenecks, but do not contain some themselves. Therefore the reasoning applies for the whole intermediate destination area and not only the part next to the virtual obstacle.

\subsection{Location and the finite extent of intermediate destination areas}

\begin{figure}[htbp]
  \center
	\includegraphics[width=0.4\textwidth]{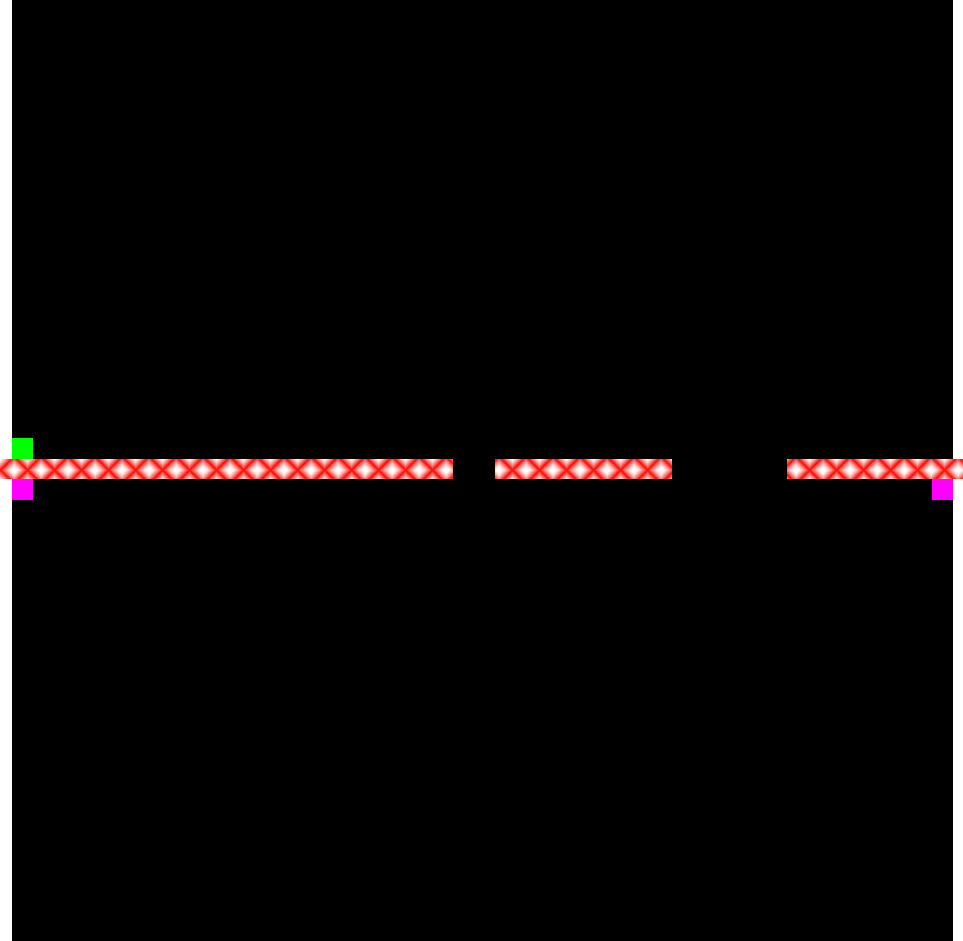} \hspace{6pt}
  \includegraphics[width=0.4\textwidth]{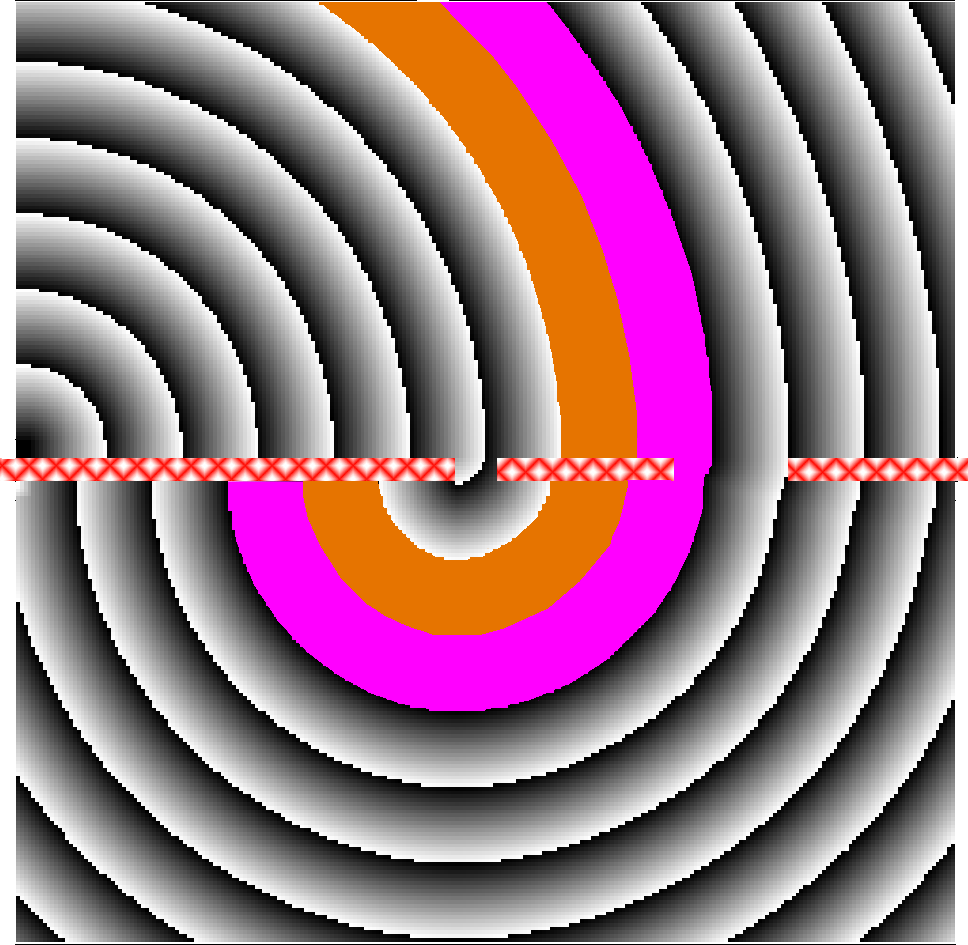} \\ 
	\caption{The figure on the left shows the geometry. Walls are colored red-white, areas to walk on black, the destination is shown green, origins magenta. The way from the left origin poses the problem, the origin on the right has been added to show that the problem does not occur for all origin positions. The figure on the right shows the result of a computation as described above, where the magenta area is used to identify the two orange areas as intermediate destinations of a routing alternative.}
	\label{fig:back1}
\end{figure}

\begin{figure}[htbp]
  \center
	\includegraphics[width=0.4\textwidth]{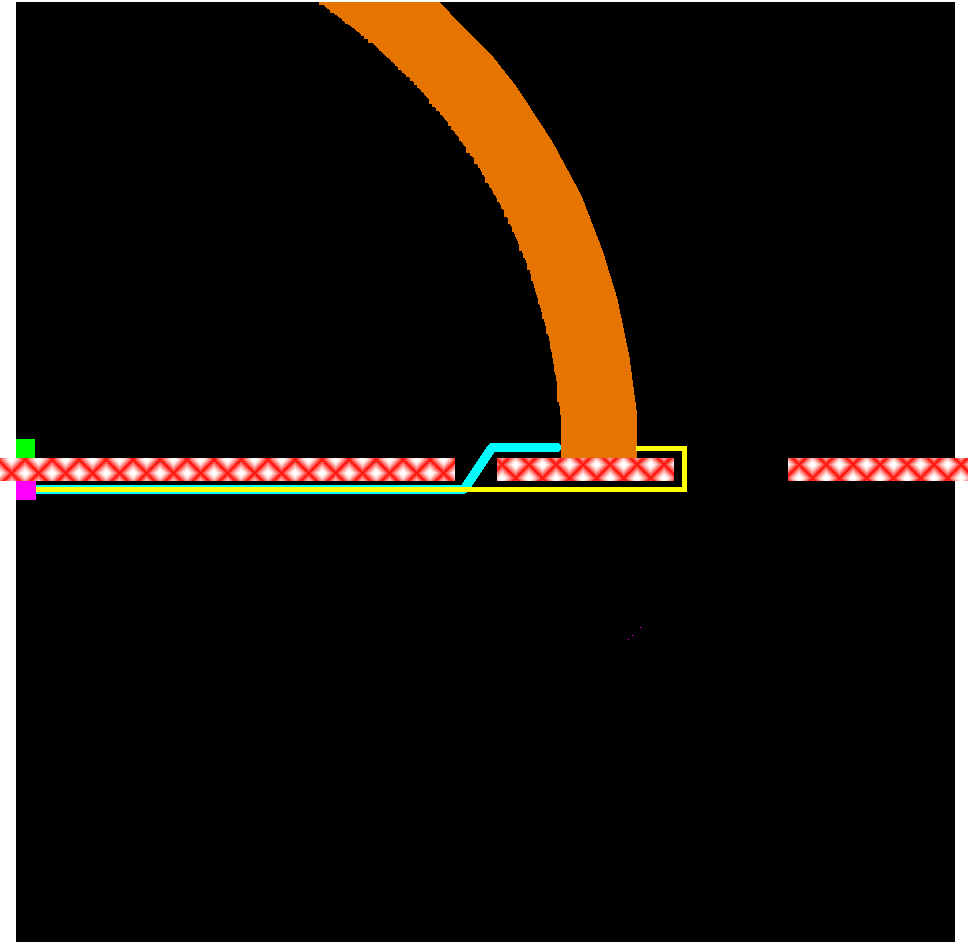} \hspace{6pt}
  \includegraphics[width=0.4\textwidth]{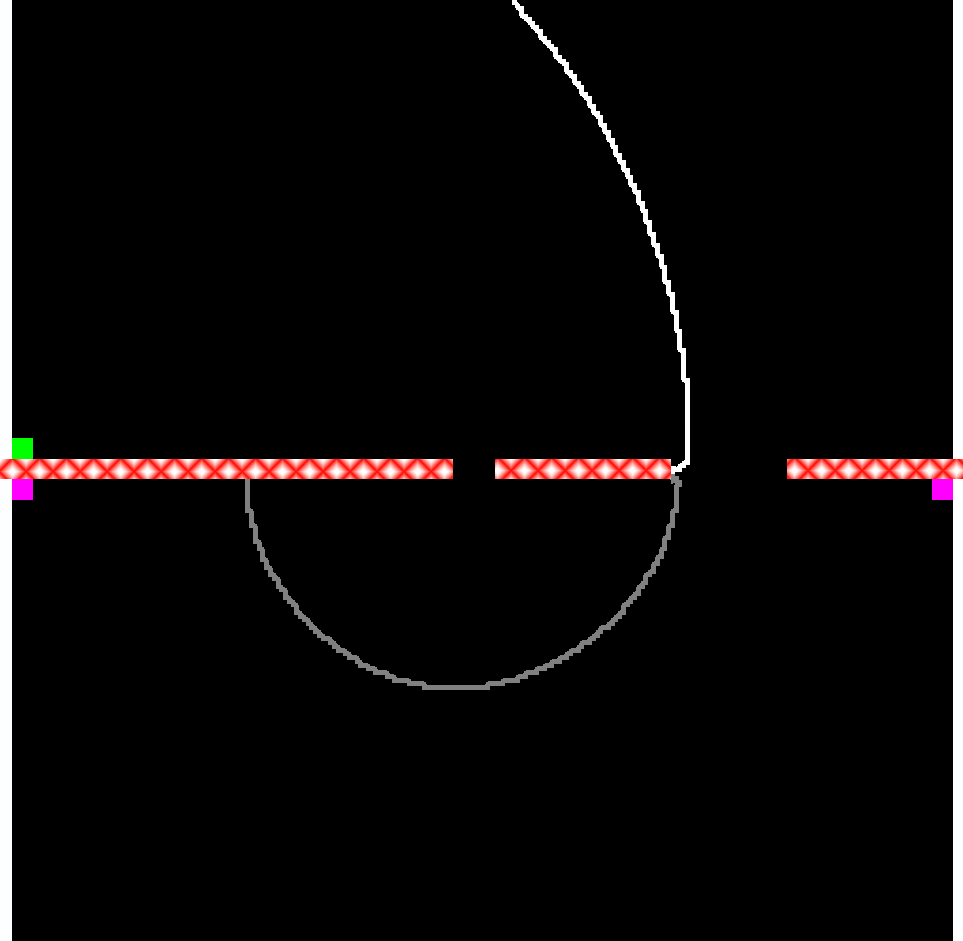} \\ 
	\caption{From figure \ref{fig:back1} (right side) here on the left side only the intermediate destination area of the detouring route is shown. In addition there are two paths shown from the origin to the front (yellow line) and to the back side (cyan line) of the intermediate destination area. Obviously the yellow path is the one which a pedestrian ought to walk (approximately). However, it can clearly be seen that the cyan path is the shorter one. Thus simulated pedestrians would walk to the wrong side of the intermediate destination area, turn around and proceed to the destination. Needless to say that this is highly unrealistic. The problem cannot be solved by only making the intermediate destination area narrower as the cyan line would even be shorter if the intermediate destination would only consist of its front edge. The core of the problem is that the front edge is not placed as far upstream as possible. The most possible upstream position is shown in the right figure. It is there where the upstream edges (one shown white, one shown gray) of the two intermediate destination areas just meet -- or more precise: just not yet meet. That this is really the most possible upstream position can easily be understood: if one would push them further upstream they would cross and a pedestrian arriving at a certain point would be closer to the destination via the other route.}
	\label{fig:back2}
\end{figure}

Finally we have to come back to the previously mentioned problem that there are geometries, where the value for parameter $d$ in principle has to be very small. Figures \ref{fig:back1} and \ref{fig:back2} shows such an example. In fact it is not just an arbitrary example, but one of a class where the problem is most pronounced. This is the case when one or more of the detouring routes includes a u-turn. Put in different words: scenarios are problematic which include sharp turns respectively strong zig-zagging.

This problem is for three reasons not as severe for the method presented in this contribution as it might appear at first. The first reason is one of relevance: typically structures where pedestrians move are built such that pedestrian paths are curved but principally forward-bound, i.e. looking at Figure \ref{fig:back1} one has rather to deal with situations similar to movement from the origin on the {\em right} side to the destination than from the origin on the left side. In these cases there is no problem no matter how large the value of parameter $d$ might be.

The second argument is that albeit in this contribution for purely illustrative reasons we have determined the extend -- to be precise: the depth -- of the intermediate destination areas from the value of parameter $d$ this is anything but required. The depth of the intermediate destination areas can be set independent of the value of parameter $d$. The downstream edge does not even have to be bended in line with equi-distance lines, but could be a straight line. In many cases the problem will vanish if one chooses a depth for the intermediate destination areas which is just large enough so a pedestrian cannot jump over it within one simulation time step. In the extreme case of a simulation time step of one second and a pedestrian speed of 3 m/s the required depth would be 3 m. More typical would be a simulation time step of 0.1 seconds and a speed of 1.5 m/s resulting in a minimum required depth of 15 cm.

Reducing the depth of the intermediate destination area will not help for the example of Figure \ref{fig:back1} as already the upstream edge is located too far downstream. However, these cases can be found and can also be fixed. One first has to check if every point on the edge of the starting area -- or more general: the next upstream intermediate destination area -- is closer to the upstream edge of the particular intermediate destination area. If this is not the case one has to shift the intermediate destination area upstream. So the third argument is: in those rare cases when the problem occurs it can reliably be found and fixed.

Still the issue might re-enforce doubts on the benefit of parameter $d$. To accept the additional effort described in this subsection that has to be done as a consequence of this approach, we have to recall that with the function related to parameter $d$ we can determine the order of size of obstacles which trigger an explicit routing alternative. In other words: it is in this way possible to define for agents (pedestrians) moving not on links but freely in 2D which two routes are sufficiently distinct. The modulo approach with parameter $d$ is about the existence or non-existence of intermediate destinations. This is highly beneficial. The side-effects concerning the position of the intermediate destinations compared to that are small and can furthermore be fixed.

\section{A small example} \label{sec:example}
\begin{figure}[htbp]
  \center
	\includegraphics[width=0.612\textwidth]{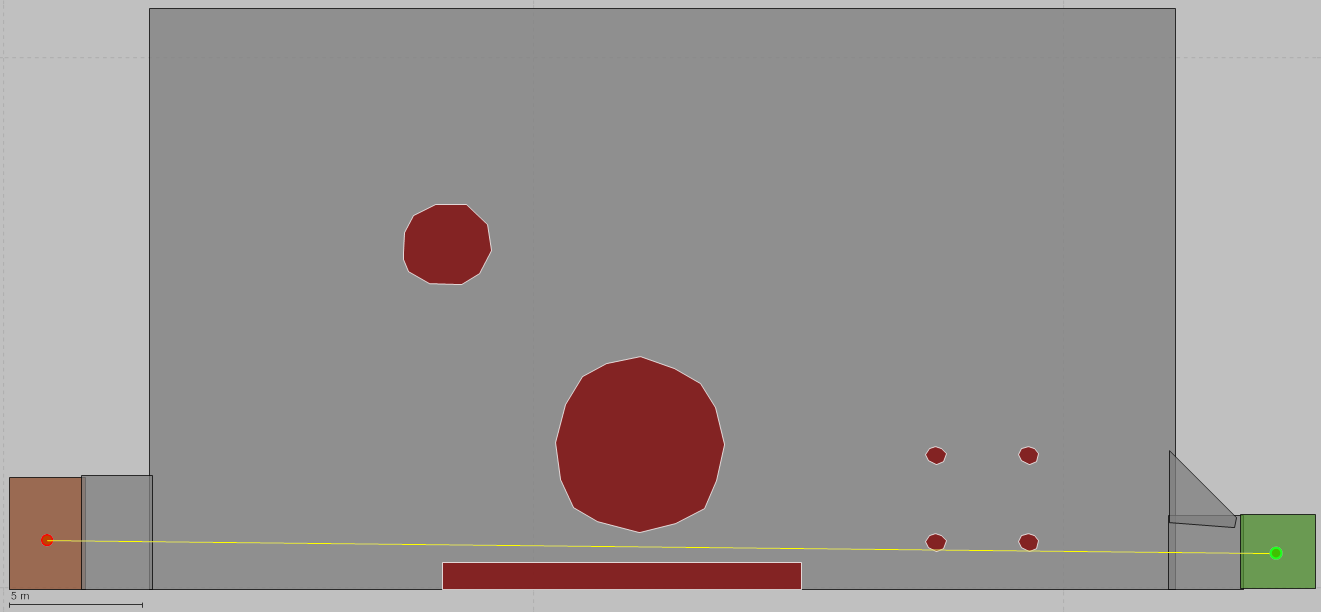}
	\caption{The example scenario. Walking areas are shown dark gray with a black edge, non-walking light gray, obstacles (i.e. also non-walking) dark red with a white edge. The origin is to the left (light red), the destination to the right (green). Of course both are walking areas as well.}
	\label{fig:Bsp0}
\end{figure}

As this contribution focuses on the construction of the intermediate destination areas and the route alternatives we only present a minimal example of application with an assignment calculation. Figure \ref{fig:Bsp0} shows the geometry of the example. There are two large obstacles which have just an intermediate size that they can be seen as obstacles in one large room or as walls separating two different rooms. In addition there are four smaller obstacles which we want to have handled by the operational simulation model of pedestrian dynamics. 

\begin{figure}[htbp]
  \center
	\includegraphics[width=0.612\textwidth]{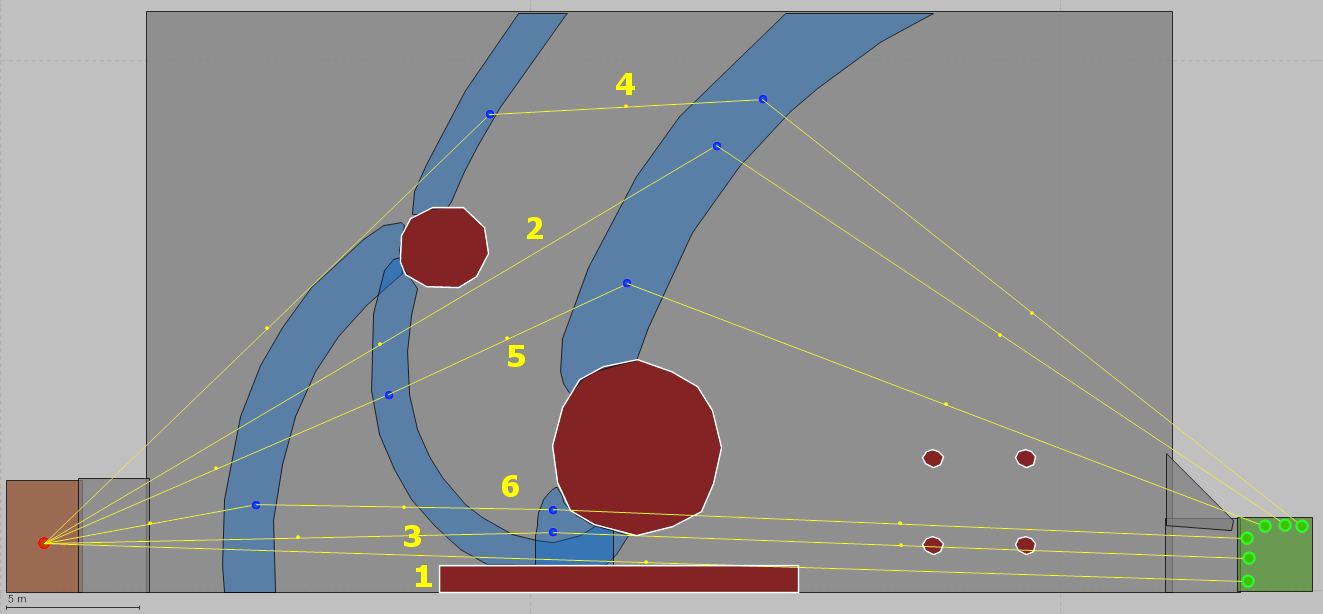}
	\caption{Resulting intermediate destinations and routes, plus IDs to identify routes in the remainder. The four small columns to the right do not create route alternatives, only the two larger obstacles in the center / to the right do.}
	\label{fig:Bsp0routes}
\end{figure}

The algorithm proposed above gives the routes and intermediate destinations as shown in figure \ref{fig:Bsp0routes}. 

It can easily be seen in figure \ref{fig:Bsp0routes} that for the origin area as existing in this scenario routes 1, 3, and 6 as well as routes 2 and 5 in general imply the same routing for pedestrians\footnote{However, note that if the origin area was located somewhere else, for example in the upper left corner this would be different.}. Transferring a concept from road traffic network assignment we can say they have a commonality factor of 1 \cite{cascetta1996modified}, although a strict and general definition of the commonality factor for pedestrian routes will not be given here. Whereas the problem of routes with a high commonality factor is known from and relevant in vehicular traffic assignment, routes with a commonality factor of 1 do not exist there, respectively the problem is irrelevant. Such routes in road traffic could easily be spotted and eliminated as the data defining them would typically be identical (the same set of links). 

With the line of argumentation pointed out in subsection \ref{sec:MO} one could drop routes 1, 2, and 3. So in this example all routes with a commonality factor of 1 could be dropped. For this initial example we will, however, also investigate the case with six routes to get an idea of the effect of routes with a commonality factor of 1.

First we will present the results of assignments iterations with only the routes 1, 2, and 4 respectively 4, 5, and 6. In these two cases no two routes do identical routing. The former case (1, 2, 4) includes routes with as few intermediate destinations as possible (i.e. much ``freedom'' for pedestrians or better said for the operational simulation model of pedestrian dynamics), the latter case includes routes with a maximum of intermediate destinations (i.e. as few freedom as possible).

In each computation the inflow on the origin area is on average 12,000 pedestrians per hour. This is over the capacity of the shortest path, but well below the total capacity of all paths. The relevant time interval is t=300 .. 600 seconds after simulation start. In each iteration only one simulation run has been carried out. The travel times of all pedestrians who arrive in this interval are considered for the assignment in the next iteration. The iteration process was continued until the difference between the longest and the shortest average route travel time was 0.5 seconds or less or until the shifted route choice ratio was smaller than 0.0005. These are rather strict conditions for termination. Yet for this example we are interested in all effects that occur in the course of many iterations.

The computation of the route choice ratios from the travel times was deliberately done rather simple. Between the pair with the longest and the shortest travel times ($t_{Max}$ and $t_{Min}$) the following probability $\Delta p$ was shifted (obviously from the route with the longer to the route with the shorter travel time):
\begin{equation}
\Delta p = \alpha \left(\frac{t_{Max} - t_{Min}}{t_{Max} + t_{Min}}\right)^\delta
\end{equation}
where $\alpha$ is a general sensitivity factor which was chosen to be $\alpha=0.1$ in all computations and $\delta$ is a dynamic adaptation factor which usually was $\delta=1$, but was decreased when the routes with the longest and the shortest travel time were identical in subsequent iterations and which was increased when they exchanged roles in subsequent iterations.

\begin{figure}[htbp]
  \center
	\includegraphics[width=0.612\textwidth]{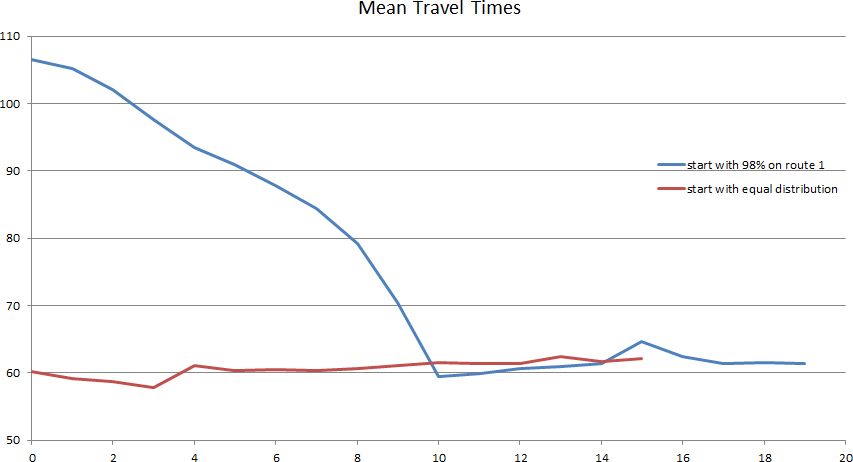} \\ \vspace{12pt}
	\includegraphics[width=0.612\textwidth]{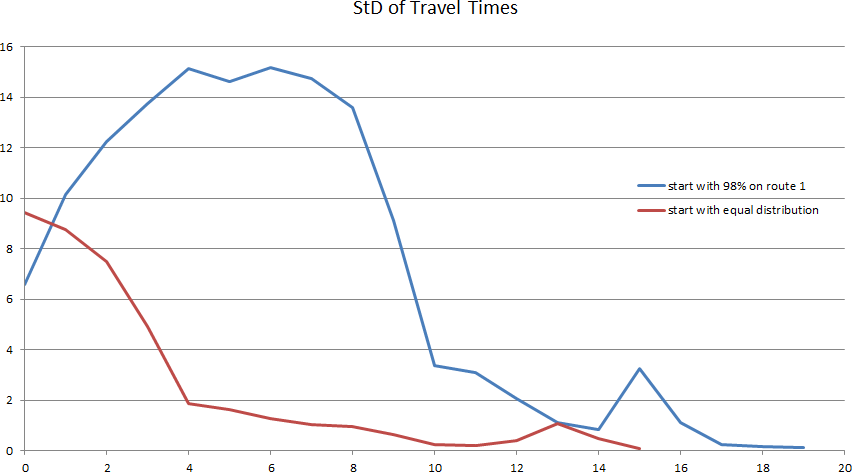}
	\caption{Weighted average travel time and weighted standard deviations of travel times -- both in [s] -- in the course of iterations when only routes 1, 2, and 4 are utilized. As a consequence of having set parameter $d$ to a value of $d=2$ m the two lines show the results when (blue line) in the first iteration 98\% of pedestrians choose route 1 (and the other two routes are used by 1\% each) and (red line) when the initial distribution is equal.}
	\label{fig:Bsp0results124a}
\end{figure}

\begin{figure}[htbp]
  \center
	\includegraphics[width=0.612\textwidth]{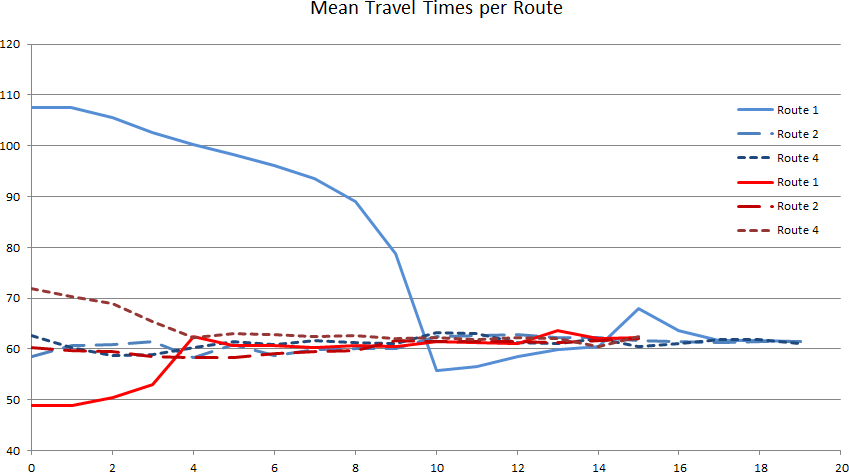} \\ \vspace{12pt}
	\includegraphics[width=0.612\textwidth]{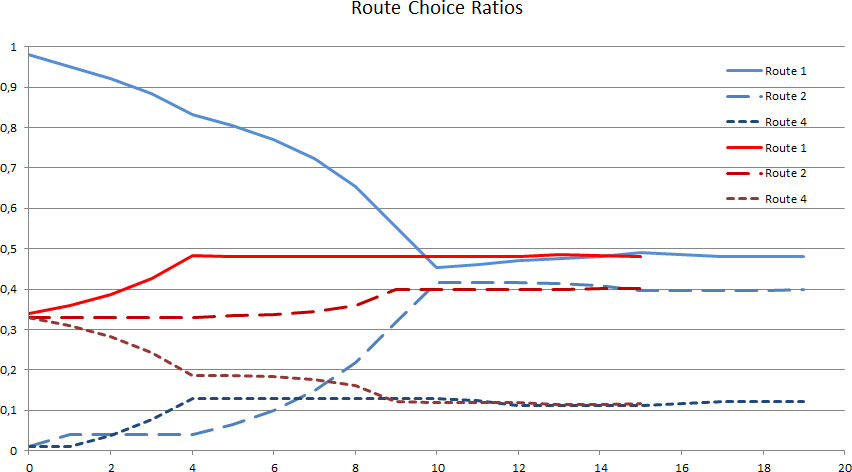}
	\caption{Travel times on each of the routes and route choice ratios vs. iteration number in the case when only routes 1, 2, and 4 are utilized. The case with 98\% on route 1 in the first iteration is shown blue, equal distribution in the first iteration red.}
	\label{fig:Bsp0results124b}
\end{figure}

Figures \ref{fig:Bsp0results124a} and \ref{fig:Bsp0results124b} show the results for two different initial distributions when only routes 1, 2, and 4 are utilized. When nearly all pedestrians are assigned to route 1 (i.e. the original route without intermediate destinations) in the first iteration the process terminates after iteration 19 with a weighted average travel time of 61.4 seconds. Beginning with equal distribution results in a travel time of 62.2 seconds after 15 iterations.

\begin{figure}[htbp]
  \center
	\includegraphics[width=0.612\textwidth]{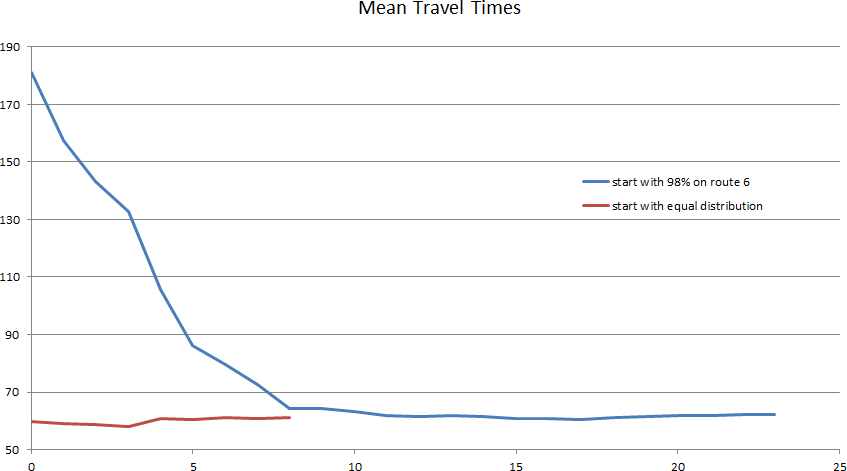} \\ \vspace{12pt}
	\includegraphics[width=0.612\textwidth]{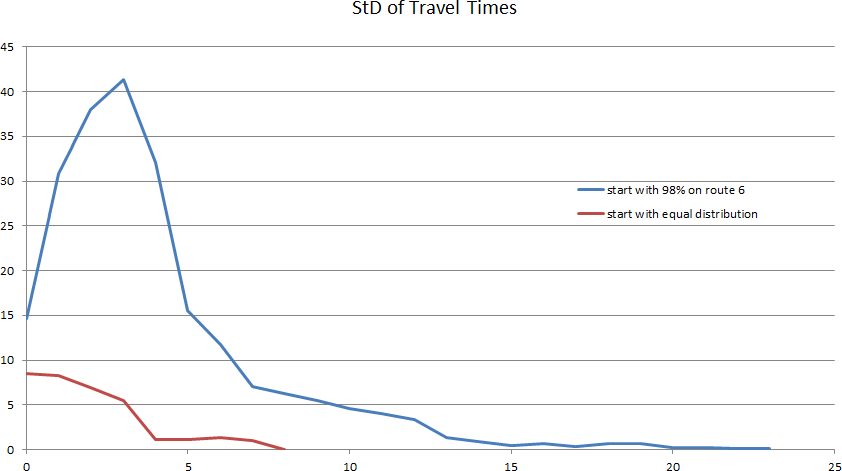}
	\caption{Weighted average travel time and weighted standard deviations of travel times -- both in [s] -- in the course of iterations when only routes 4, 5, and 6 are utilized. The two lines show the results when (blue line) in the first iteration 98\% of pedestrians choose route 1 (and the other two routes are used by 1\% each) and (red line) when the initial distribution is equal.}
	\label{fig:Bsp0results456a}
\end{figure}

\begin{figure}[htbp]
  \center
	\includegraphics[width=0.612\textwidth]{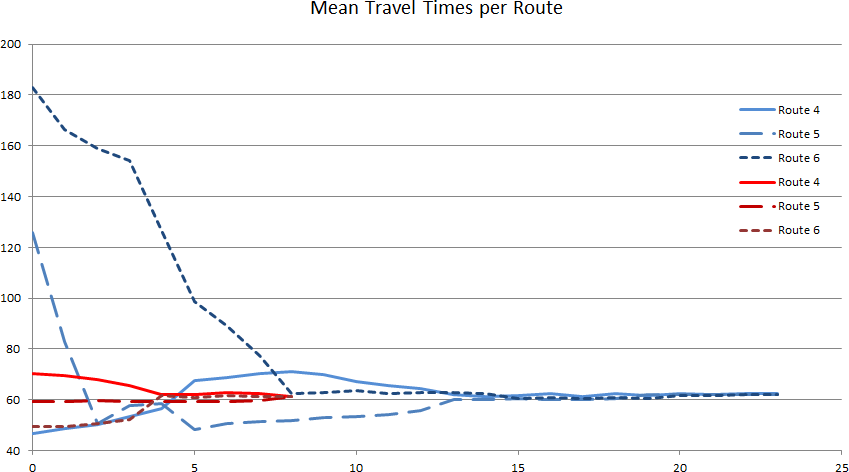} \\ \vspace{12pt}
	\includegraphics[width=0.612\textwidth]{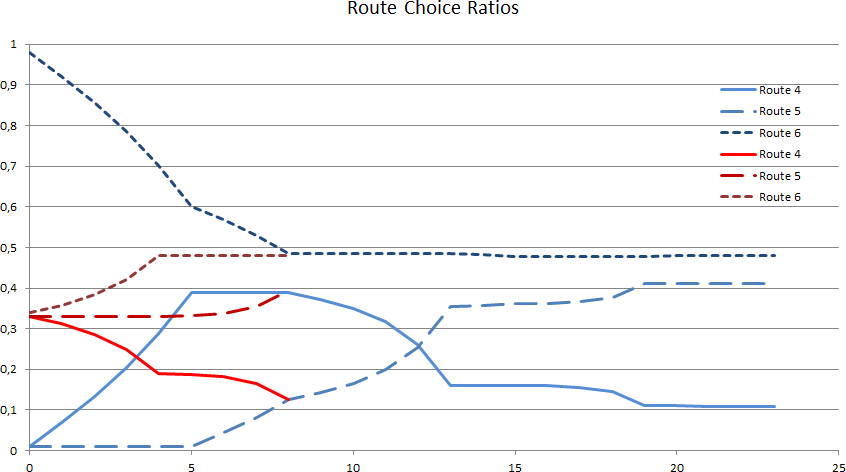}
	\caption{Travel times on each of the routes and route choice ratios vs. iteration number when only routes 4, 5, and 6 are utilized. The case with 98\% on route 1 in the first iteration is shown blue, equal distribution in the first iteration red.}
	\label{fig:Bsp0results456b}
\end{figure}

Figures \ref{fig:Bsp0results456a} and \ref{fig:Bsp0results456b} show the corresponding results when only routes 4, 5, and 6 are utilized. When nearly all pedestrians are assigned to route 6 in the first iteration the process terminates after iteration 23 with a weighted average travel time of 62.0 seconds. Beginning with equal distribution results in a travel time of 61.2 seconds after 8 iterations. 

\begin{figure}[htbp]
  \center
	\includegraphics[width=0.612\textwidth]{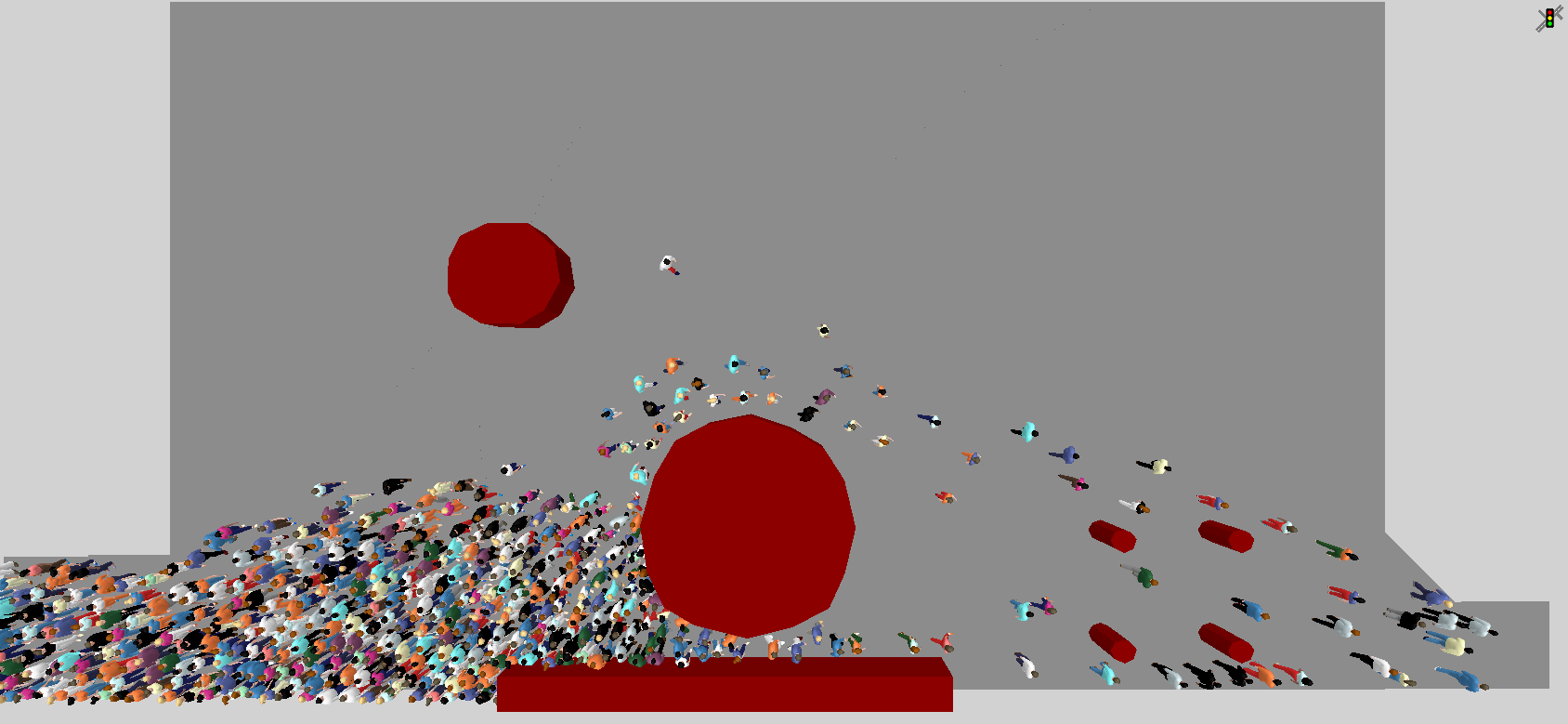} \\ \vspace{12pt}
	\includegraphics[width=0.612\textwidth]{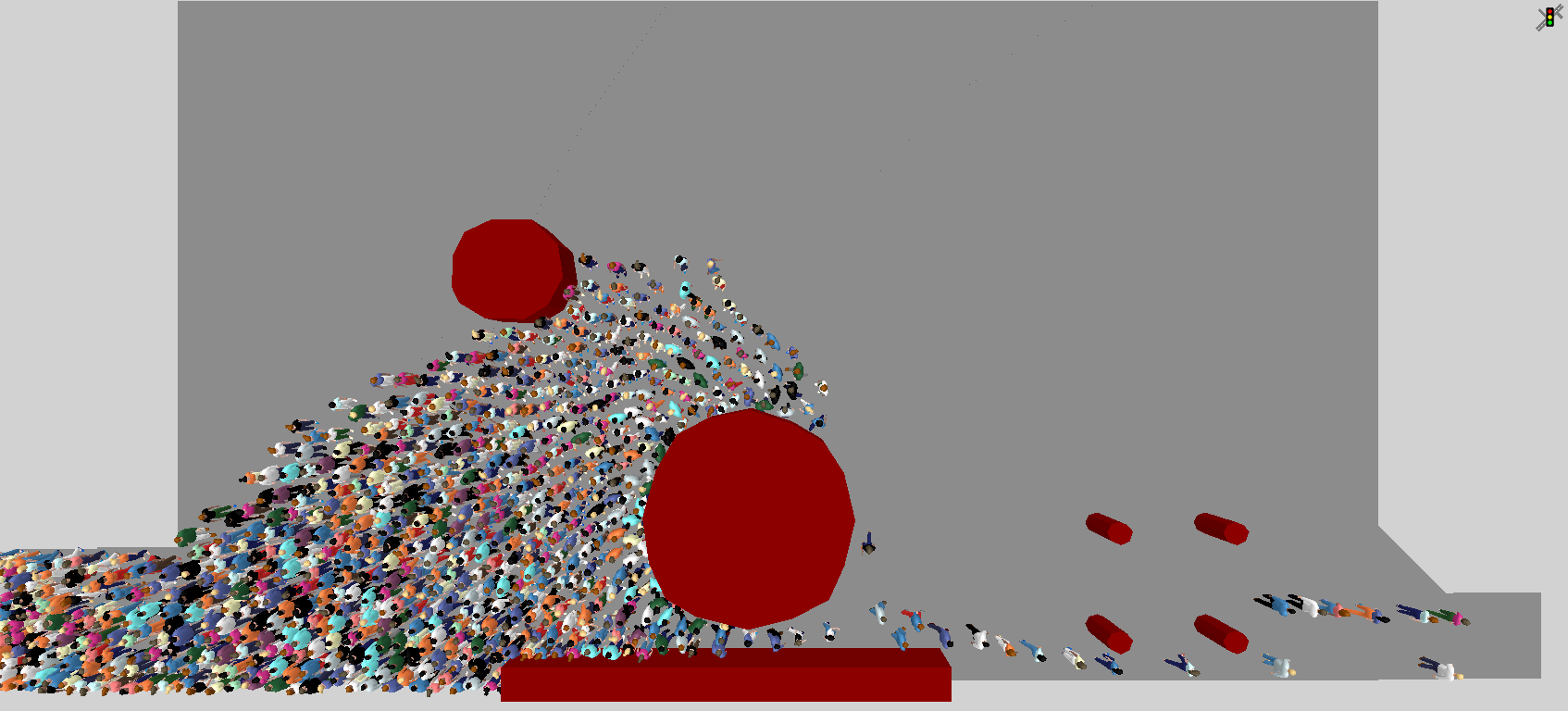}
	\caption{Still images from the first iteration step of the case with routes 1, 2, and 4 (upper figure) and 4, 5, and 6 (lower figure). Nearly all pedestrians are sent along route 1 or 6 respectively. As the demand clearly exceeds the capacity of these two routes a jam forms. In the former case at some point it is shorter (fewer meters) for pedestrians to pass the largest obstacle on the left side. I.e. although they are sent on route 1, they move on route 2, disturbing the assignment algorithm.  In the latter case the second intermediate destination of route 6 which is located between the two larger obstacles prevents this from happening for a long time and when it happens pedestrians do not walk directly to the destination, but first have to approach the back side of the second intermediate destination area of route 6 also disturbing the assignment process.}
	\label{fig:Bsp0scenes}
\end{figure}

Table \ref{tab:124vs456} summarizes the results of the cases when only three routes are utilized. It can be seen that under different conditions the results of the assignment process agree closely. This is a good sign for the quality of the algorithm that computes the intermediate destination areas: obviously it does not make a difference if pedestrians are sent via route 1 or 6 and respectively route 2 or 5. Figure \ref{fig:Bsp0scenes} shows a comparison of the 1, 2, 4 and the 4, 5, 6 routes computations during the first iteration step when nearly all pedestrians are sent via route 1 respectively 6.

To get an estimation what the remaining differences for mean travel times in table \ref{tab:124vs456} imply we have done four more computations with different initialization values for the random number generator of the pedestrian simulation for the ``4, 5, 6, equal'' case. The minimum average travel time of the overall ten computations was 61.2 s, the maximum 65.6 s and the average 63.7 s with a standard deviation of 1.3 s. This shows that the results given in table \ref{tab:124vs456}  are not more disperse than for one and the same case with only different random numbers.

\begin{table}[htbp]
	\center
	\begin{tabular}{|l|cc|cc|cc|c|} \hline
initial      &\multicolumn{2}{|c|}{Route 1}&\multicolumn{2}{|c|}{Route 2}&\multicolumn{2}{|c|}{Route 4}&Mean\\
distribution &r.c.r.	&t.t.	               & r.c.r.	 &t.t.	             &r.c.r.	&t.t.                &t.t.\\ \hline
98\% route 1 &0.480	  &61.5	               &0.399	   &61.4	             &0.121	  &61.1                &61.4\\
equal dist.  &0.482	  &62.2	               &0.402	   &62.1	             &0.117	  &62.4                &62.2\\ \hline
		         &\multicolumn{2}{|c|}{Route 6}&\multicolumn{2}{|c|}{Route 5}&\multicolumn{2}{|c|}{Route 4}&Mean\\
	           &r.c.r.	&t.t.	               & r.c.r.	 &t.t.	             &r.c.r.	&t.t.                &t.t.\\ \hline
98\% Route 6 &0.480	  &62.0	               & 0.411	 &62.0	             &0.109	  &62.4                &62.0\\
equal dist.  &0.480	  &61.2	               & 0.394	 &61.2	             &0.126	  &61.3                &61.2\\ \hline
	\end{tabular}
	\caption{Summary of results when only three routes are utilized (1, 2, and 4 or 4, 5, and 6): the route choice ratios (r.c.r.) and travel times (t.t.) per route and in the last column the weighted average travel time. As route 6 corresponds to route 1 and route 5 corresponds to route 2 in the lower half of the table the sequence of columns does not follow the sequence of route IDs.}
	\label{tab:124vs456}
\end{table}

If the iteration process is done utilizing all six routes, there is the additional difficulty that three routes (1, 3, and 6) and two routes (2 and 5) effectively contain identical navigation information as long as no phenomenons as shown in figures \ref{fig:Bsp0scenes} occur. We have started the assignment process three times with different initial assignment: a) 95\% on route 1 and 1\% on each of the other routes; b) 34\% resp. 33\% on routes 1, 3, and 6 and 1\% on each of the three remaining routes; c) equal distribution for all routes. Table \ref{tab:123456a} shows the results 

\begin{table}[htbp]
	\center
	\begin{tabular}{|l|cccccc|c|} \hline
init. distr. & R 1  & R 2	  & R 3 &	R 4	  & R 5	  & R 6	 & Mean \\ \hline
             &\multicolumn{7}{|c|}{Travel Times [s]} \\ \hline
a)           & 63,2	& 63,4	& -	  & 63,8	& 63,2	& -	   & 63,24 \\
b)	         & 61,4	& 61,5	& -	  & 61,4	& 61,6	& 61,1 & 61,43 \\
c)	         & 63,3	& 63,5  & -	  & 63,5	& 63,4	& 63,4 & 63,39 \\ \hline
             &\multicolumn{6}{|c|}{Route Choice Ratios}&Final Iteration \\ \hline
a)           &0,487	& 0,016 & 0   & 0,062	& 0,436	&0     &160 \\
b)	         &0,435	& 0,320 & 0   & 0,105	& 0,085	&0,056 &152 \\
c)	         &0,439	& 0,246 & 0   & 0,065	& 0,193	&0,057 &175 \\ \hline
	\end{tabular}
	\caption{Results when all six routes are utilized. The table shows the average travel times per route and the route loads of the final iteration (the number which is shown in the lower right column).}
	\label{tab:123456a}
\end{table}

\begin{figure}[htbp]
  \center
	\includegraphics[width=0.612\textwidth]{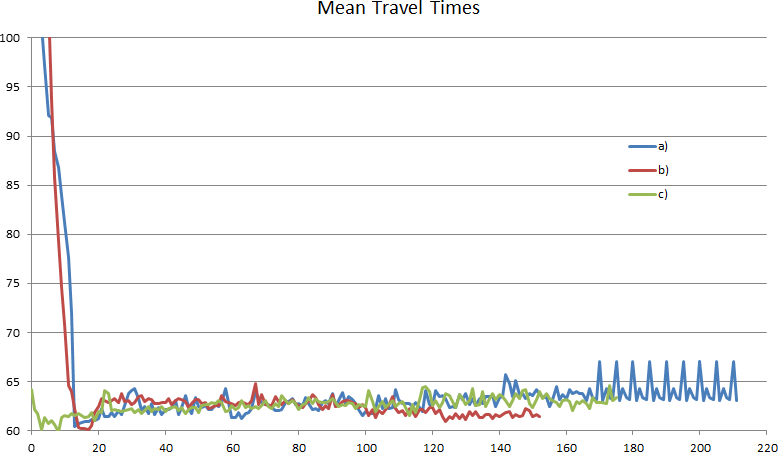} \\ \vspace{12pt}
	\includegraphics[width=0.612\textwidth]{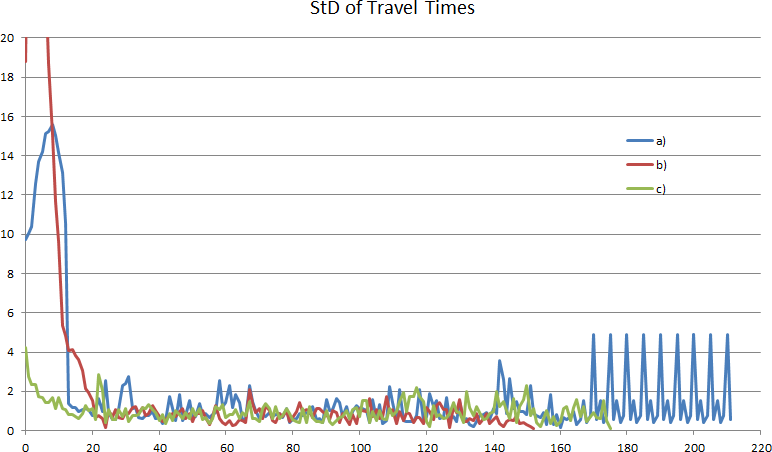}
	\caption{Weighted average travel time and weighted standard deviations of travel times -- both in [s] -- in the course of iterations when all routes are utilized. For reasons of clarity the diagrams are cut and do not show all values. For initial conditions b) the maximum mean value of travel times was 175.2 s and the maximum standard deviation 35.4 s.}	\label{fig:Bsp0results123456tt}
\end{figure}

Reading table \ref{tab:123456a} carefully one sees that for initial conditions a) the maximum travel time difference is 0.6 and not 0.5 or smaller as required for termination of the process. This is not because there were no more changes to the route choice ratios. Instead the reason is that for initial conditions a) starting at about iteration 165 the assignment process entered a loop in which the termination condition was not fulfilled and thus the iteration process would continue forever. An effect of that phenomenon can be seen in figure \ref{fig:Bsp0results123456tt}. This is a problem of a) having strict termination conditions\footnote{If the termination condition had been 1.0 seconds then termination had occurred after at most 21 iterations instead of something between 160 and infinity.} and b) the assignment method which we deliberately chose to be simple for this initial work on the topic. To fix this one could do more than one simulation run per iteration and calculate rout choice ratios based on the averages of these runs. Another fix would be to elaborate the assignment method by for example considering not only the last iteration step when calculating new route choice ratios but also the second last or even more to the extreme of considering all iterations. Here we do not elaborate the assignment method, but simply chose to take that iteration step as result for initial conditions which a) which was as close as possible to the termination conditions.

\begin{figure}[htbp]
  \center
	\includegraphics[width=0.612\textwidth]{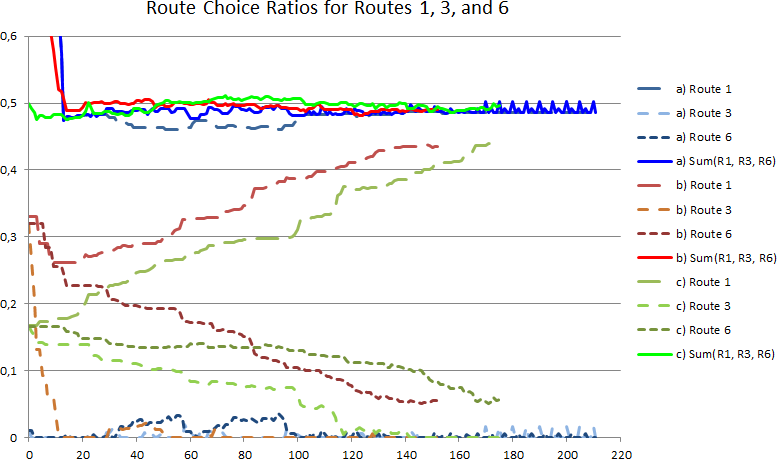} \\ \vspace{12pt}
	\includegraphics[width=0.612\textwidth]{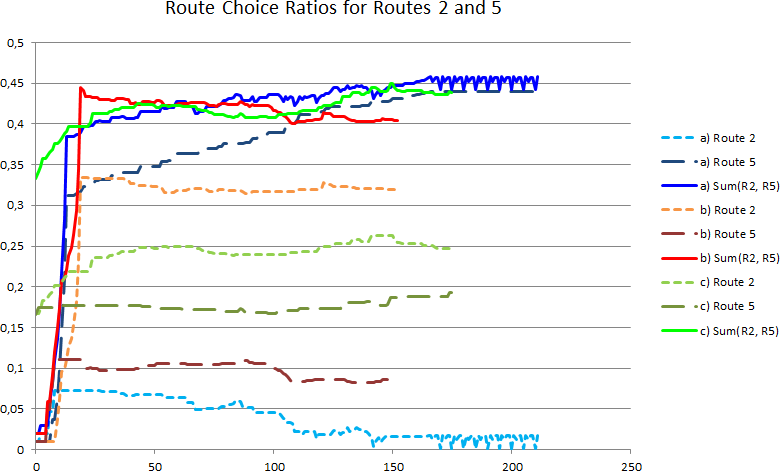}
	\caption{Route choice ratios in the course of iterations for the three initial conditions a, b, and c.}	
	\label{fig:Bsp0results123456rcr}
\end{figure}

Interesting in table \ref{tab:123456a} is also that -- while the sum of both is similar -- the route choice ratios for routes 2 and 5 show large differences with respect to the initial conditions: for initial condition a) route 5 is used heavier, while for b) and c) route 2 is made more use of. This is different for routes 1, 3, and 6 for which the route choice ratios end up comparable if they are compared across all three initial conditions. Looking in figure \ref{fig:Bsp0results123456rcr} at the evolution of the route choice ratios in the course of iterations one can recognize another difference: while the sums of route choice ratios (1, 3, and 6 as well as 2 and 5) both quickly are very stable this is only the case for the individual routes 2 and 5 while the ratios for particularly the routes 3 and 6 keep changing for many iterations for at least two of the three initial conditions.

That there is no unique solution for the route choice ratios of routes 2 and 5 when the iteration process terminates does not indicate a problem. It rather confirms that in fact and as intended the intermediate destinations which were introduced have no artificial impact on local walking behavior. If they had, they would most probably also have an impact on travel times and if that was the case the travel times for routes 2 and 5 would not be identical. Only if they are identical there is no unique solution for their choice ratios. Put in different words: The intermediate destination which is part of route 5, but not of route 2 does neither de- nor increase the travel time of pedestrians on route 5 compared to pedestrians on route 2. If it would do so route 5 would always be a better or always be a worse choice than route 2. Than route 5 over the iterations would by and by gain or lose choice ratio compared to route 2 and this in all three cases of initial conditions. That this is not the case is a good indication that the intermediate destination area which is part of route 5 but not of route 2 is shaped such that it does not introduce artifacts to the movement of pedestrians. Strictly speaking we have this indication only for this one single intermediate destination area, but one can be confident that this is as well the case for all other intermediate destination areas.

Does this on the contrary imply that on routes 3 and 6 there is a delay due to locally different movement behavior introduced by the intermediate destinations? This cannot be the case, as route 6 for two of the initial conditions has a higher load than route 3, although it has an additional intermediate destination. The cause is rather the phenomenon shown in figure \ref{fig:Bsp0scenes}. Particularly the intermediate destination between the rectangle wall and the largest circular obstacle -- which is part of routes 3 and 6, but not 1 -- imposes reduced flexibility for all pedestrians who have to reach for it. This can increase the average travel time on routes 3 and 6 because of a few pedestrians who experience extended delay times instead of passing the largest obstacle to the left, effectively walking on route 2 or 5. Still, their travel time will be recorded as belonging to route 1. If a pedestrian at the very same spot had been assigned to walk route 3 or 6 instead of route 1, he or she had not turned around (proceeding effectively on route 2 or 5), but had waited long to walk the intended path. Thus in this high demand situation average travel times on routes 1, 3, and 6 are different. This is an indication that generally in a case with a number of routes with a commonality factor of 1 it might be more consequent to keep the route with most intermediate destinations and drop the others.

\section{Summary and Conclusions}
We have introduced a method to generate the geometry of intermediate destinations and routes utilizing these in a two dimensional movement (walking) environment. Simulated pedestrians can follow the routes intermediate destination by intermediate destination from their origin to their destination. The shape of the intermediate destination is such that the pedestrians do not make turns when they reach an intermediate destination. This is because the intermediate destination is shaped along equi-distance lines of the next downstream (intermediate) destination. Thus someone following the animation of a simulation would not realize where intermediate destinations are located but just see pedestrians utilize many different instead of just the spatially shortest route. 

For one example the resulting route set was applied in an assignment computation: a travel-time based user equilibrium was found by iterative simulation. The result was similar for different initial conditions. The example showed that the method works as intended.

Future work includes finding a method to eliminate the remaining routes with a commonality factor of 1 and the development of a more efficient assignment method specifically for pedestrian simulation. In the latter task one has to consider that here we do not start with a node-link network as it is the case in road traffic assignment methods where the available routes are typically computed with a classical shortest path search. Instead we already set out with a set of routes, but lack the node-link network. This is one of the reasons for another important fact to consider: the capacities on each route are unknown which in turn is one of the reasons why no differentiable volume-delay functions are available. From this follows that there is no other way to get information on travel times than to carry out the full simulation which is expensive in terms of computation time. Thus it can be concluded that assignment methods are best suited which first adapt the route choice ratios of all routes in the network and only than carry proceed with another iteration step, i.e. simulation run.

\begin{appendix}
\section{Definitions}
{\em Microscopic simulation} is understood here as an algorithm in which the movement of individual entities is computed according to a given scheme. For our purpose this scheme is defined in some way such that the entities move locally similar as pedestrians move in reality (simulation of pedestrians). The interaction between entities is at least in parts computed individually (as opposed to interaction of an individual with a mean field).

{\em Simple polygon} is understood in this contribution not only as the line segments defining the polygon, but also as the enclosed region.

It is assumed that areas in which pedestrians walk as well as obstacles which exclude access can be defined with a principally unlimited number of simple polygons. The total walking area is the unification of all polygons defining walking areas subtracting the unification of all polygons defining obstacles. In other words: if one or more polygons defining walking space and one or more polygons defining an obstacle overlap the overlap area cannot be accessed by pedestrians.

It is assumed that the space in which pedestrians move can be modeled two-dimensionally.  This implies that it is neglected that in reality in a projection along the z-axis a pedestrian and an obstacle may overlap. 

Stairwells and multiple floors could be connected piecewise in this way, i.e. the third spatial dimension can be approximated for all planning needs with a number of distinct sets of two-dimensional planes. However, higher floors and stairwells are not considered further in this paper.

The {\em origin} of a particular pedestrian $i$ is the coordinate $(x_i(T_{o,i})/y_i(T_{o,i}))$ at which he is set into the simulation at time $t=T_{o,i}$. This coordinate can be one of a set which is as well a simple polygon called {\em origin area}. Different pedestrians will be set into the simulation at different times and at coordinates within the origin area.

Concerning the assignment computation aspect in this paper: the origin time $T_o$ is an extrinsic parameter for all pedestrians, i.e. demand is inelastic.

The destination is as well modeled as a simple polygon called {\em destination area}. The actual {\em destination} $(x_i(T_d)/y_i(T_d))$ of a particular pedestrian $i$ is a coordinate within the destination area. A pedestrian is considered to have arrived at his destination when he is for the first time (time $t=T_{d,i}$) located on a coordinate which is element of the destination area. This implies that each part of the destination area\footnote{respectively each part of the edge of the destination area} is of equal value for a pedestrian. Which coordinate of the destination area the pedestrian is heading for is not a property of the destination area.

An {\em intermediate destination area} is a simple polygon to which pedestrians are heading for before they are walking towards the destination area. Therefore intermediate destination areas have a very similar purpose as destination areas and they have the same effect on pedestrians except that pedestrians are not taken out of the simulation when they have arrived at a destination area. Instead they continue to walk to the next destination area or to their (final) destination area.

A {\em route} is a sequence of simple polygons which begins with an origin area, then has an arbitrary number of intermediate destination areas and ends with a destination area. Contrary to trajectories routes are understood to be extrinsic parameters for the operational simulation model, i.e. routes are input data. 

To be distinguished from a route has to be a {\em trajectory}. A trajectory of a pedestrian is defined as a curve in space which is parametrized with time. In a microscopic simulation which progresses in constant, discrete time steps -- as is the case here -- a trajectory of pedestrian $i$ is a set of coordinates $(x_i(t)/y_i(t))$ with $T_{o,i}\leq t \leq T_{d,i}$. This implies that a route -- or better the set of all routes -- is input data for the simulation, while trajectories are output data.

For completeness: a {\em path} is a trajectory stripped of its time aspect and relation to a specific pedestrian.

{\em Route decisions} are objects which hold two or more routes and a probability for each route that it is chosen by a pedestrian. When a pedestrian is set into the simulation a {\em route decision} is carried out for him. This means that among a defined set of routes which usually is different for each origin area one is chosen probabilistically and assigned to the pedestrian. Like routes route decisions are input data for the operational level of the simulation. The route decision probabilities are numbers which are external input data. The values are not affected by anything which happens within a simulation run, they are in principle fixed over time. However, there can be different time intervals in which the route decision probabilities have different values. Concerning the assignment computation aspect in this paper: if there is more than one time interval for the route decision probabilities, i.e. if the route decision probabilities are not constant throughout the simulation, the assignment is a {\em dynamic assignment}. As it is a micro-simulation the time intervals in principle can be arbitrarily small which is an advantage if demand varies on short time scales. However, for usage with an iterated assignment method the time intervals need to be sufficiently long to produce a meaningful statistics (average) of travel times. If this is not given one cannot expect convergence of the assignment process.

An {\em OD matrix} is the integrated representation of demand, routes, and route decisions. The lines of the matrix comprise of the origin areas, the columns of the destination areas. Each entry of the matrix contains the absolute demand volume that within a given time interval is meant to move from origin area to destination area.

\end{appendix}

%
\nocite{_Ency2009book,_ACRI2010}
\bibliographystyle{utphys2011}
\bibliography{PedAssignment}
%
\end{document}